# Current laboratory performance of starlight suppression systems, and potential pathways to desired Habitable Worlds Observatory exoplanet science capabilities

**Bertrand Mennesson,[a,*] Ruslan Belikov,[b] Emiel Por,[c] Eugene Serabyn,[a] Garreth Ruane,[a] A.J. Eldorado Riggs,[a] Dan Sirbu,[b] Laurent Pueyo,[c] Remi Soummer,[c] Jeremy Kasdin,[d] Stuart Shaklan,[a] Byoung-Joon Seo,[a] Christopher Stark,[c] Eric Cady,[a] Pin Chen,[a] Brendan Crill,[a] Kevin Fogarty,[b] Alexandra Greenbaum,[f] Olivier Guyon,[g] Roser Juanola-Parramon,[e] Brian Kern,[a] John Krist,[a] Bruce Macintosh,[h] David Marx,[a] Dimitri Mawet,[a,i] Camilo Mejia Prada,[a] Rhonda Morgan,[a] Bijan Nemati,[j] Leonid Pogorelyuk,[k] Susan Redmond,[i] Sara Seager,[l] Nicholas Siegler,[a] Karl Stapelfeldt,[a] Sarah Steiger,[c] John Trauger,[a] James K. Wallace,[a] Marie Ygouf,[a] & Neil Zimmerman[e]**

[a]Jet Propulsion Laboratory, California Institute of Technology, 4800 Oak Grove Drive, Pasadena CA 91109, USA
[b]NASA Ames Research Center, Moffett Field, CA, USA 94040
[c]Space Telescope Science Institute, 3700 San Martin Drive, Baltimore, MD, USA
[d]Dept. of Mechanical and Aerospace Engineering Princeton University, Princeton, New Jersey 08544, USA
[e]NASA Goddard Space Flight Center, 8800 Greenbelt Rd, Greenbelt, MD 20771, USA
[f]IPAC, California Institute of Technology, 1200E. California Boulevard, Pasadena, CA 91125, USA
[g]Subaru Telescope, National Astronomical Observatory of Japan, National Institutes of Natural Sciences (NINS), 650 North A'ohoku Place, Hilo, HI 96720, USA
[h]University of California Observatories, 550 Red Hill Rd, Santa Cruz, CA 95064
[i]Department of Astronomy, Caltech, 1200E. California Boulevard, Pasadena, CA, 91125, USA
[j]Tellus 1 Scientific, LLC, 8401 Whitesburg Dr SE, Huntsville, AL 35802, USA
[k]Rensselaer Polytechnic Institute, 110 Eighth Street, Troy, NY, 12180, USA
[l]Department of Earth, Atmospheric and Planetary Sciences, Massachusetts Institute of Technology, 77 Massachusetts Avenue, Cambridge, MA, 02139, USA

**Abstract**. We summarize the current best polychromatic (~ 10 − 20 % bandwidth) contrast performance demonstrated in the laboratory by different starlight suppression approaches and systems designed to directly characterize exoplanets around nearby stars. We present results obtained by internal coronagraph and external starshade experimental testbeds using entrance apertures equivalent to off-axis or on-axis telescopes, either monolithic or segmented. For a given angular separation and spectral bandwidth, the performance of each starlight suppression system is characterized by the values of "raw" contrast (before image processing), off-axis (exoplanet) core throughput, and post-calibration contrast (the final 1-sigma detection limit of off-axis point sources, after image processing). Together, the first 2 parameters set the *minimum* exposure time required for observations of exoplanets at a given signal-to-noise, i.e., assuming perfect subtraction of background residuals down to the photon noise limit. In practice, residual starlight speckle fluctuations during the exposure will not be perfectly estimated nor subtracted, resulting in a finite post-calibrated contrast and exoplanet detection limit whatever the exposure time. To place the current laboratory results in the perspective of the future Habitable Worlds Observatory (HWO) mission, we simulate visible observations of a fiducial Earth/Sun twin system at 12 pc, assuming a 6m (inscribed diameter) collecting aperture and a realistic end-to-end optical throughput. The exposure times required for broadband exoearth detection (20% bandwidth around λ=0.55 μm) and visible spectroscopic observations (R=70) are then computed assuming various levels of starlight suppression performance, including the values currently demonstrated in the laboratory. Using spectroscopic exposure time as a simple metric, our results point to key starlight suppression system design performance improvements and trades to be conducted in support of HWO's exoplanet science capabilities. These trades may be explored via numerical studies, lab experiments, as well as high contrast space-based observations and demonstrations.

*Bertrand Mennesson**, E-mail: Bertrand.mennesson@jpl.nasa.gov

# 1 Introduction

To enable a broad range of exciting potential missions across the electromagnetic spectrum, the 2021 US Decadal Survey in Astronomy and Astrophysics[1] recently recommended the creation of a Great Observatories mission and technology maturation program (GOMAP). It further recommended that the first mission to be advanced under this program be a large (~6m) infrared/optical/ultraviolet space telescope that will "search for biosignatures from a robust number of about ~25 habitable zone planets and be a transformative facility for general astrophysics". This prospective mission has since been called the Habitable Worlds Observatory, or HWO.

To detect and spectrally characterize small exoplanets orbiting in the habitable zones of nearby Sun-like stars, HWO must include a powerful starlight suppression system able to cancel on-axis light very efficiently – by typically 10 orders of magnitude – while simultaneously passing a significant fraction of the nearby exoplanet light. In addition, various interesting molecules and spectral features dictate that the observations be conducted over a spectral bandwidth as broad as possible. For example, the HabEx and LUVOIR pre-decadal large mission concept studies[2,3] baselined a wavelength range of 0.2 to 1.8 μm for exoplanet direct observations, enabling searches for a variety of potential bio-signatures in rocky planets' atmospheres, and placing them into context. Obtaining near-UV to near-infrared spectra of all targets would indeed be ideal, especially for exoearth characterization (e.g.,[4,5]). However, observing in the near UV and in the infrared presents well-known additional challenges, in terms of sensitivity and spatial resolution, respectively. A more limited wavelength range of ~0.45 μm to ~1 μm has also been previously considered[6-8] for the spectral characterization of directly imaged terrestrial – and gas giant – planets in reflected light. This "visible" range would still provide access to strong spectral features for a limited collection of gases ($CO_2$, $H_2O$, $CH_4$, $O_2$, and $H_2$), as well as the opportunity to identify atmospheric Rayleigh scattering. The detailed exoplanet science objectives of HWO are yet to be specified in detail. But even if the narrower 0.45–1 μm wavelength range was adopted as a minimum threshold for direct exoplanet observations with HWO, it would already represent a spectral bandwidth of 76% (defined as $\Delta\lambda/\lambda_{center}$). This would call for a starlight suppression system that can either operate over the whole spectral range at once or observe over a limited set of "reasonably broad" (here taken to mean at least 10%) individual spectral channels within that range.

Over the last 10-20 years, many laboratory testbeds have been set up worldwide[9] to take on the challenge of broadband high-contrast[i] starlight suppression to directly image and characterize Earth-like planets. They have used different approaches, operating conditions (e.g., in air vs. vacuum), and aperture types (clear vs. obscured, monolithic vs. segmented).

Starlight can be suppressed between the telescope primary mirror and the detectors by "internal coronagraphs", or between the star and the telescope by "external coronagraphs". We will follow standard practice and refer to internal coronagraphs as simply "coronagraphs", and to external coronagraphs as "starshades".

---

[i] See Sec. 2.1 for rigorous definitions of the contrast terms used in this paper.

For coronagraphs, testbeds have been developed both in air and in vacuum. Current in-air testbeds have tackled many different – and in most cases complementary – aspects of the overall technological challenge. For instance, the "Tres Haute Dynamique" (THD2, French for "Very High Contrast") testbed[10-13] of the Paris Observatory concentrated on ultra-broadband starlight suppression, and on a specific focal plane wavefront-sensing approach using a self-coherent camera. The NASA Ames Coronagraph Experiment (ACE) testbed focused on multi-star wavefront control for high contrast imaging around binary stars[14,15], as well as phase-induced apodization techniques to provide high throughput observations at small separations[16,17] in collaboration with the Subaru SCExAO team. The High-Contrast High-Resolution Spectroscopy for Segmented telescopes Testbed (HCST) located at the California Institute of Technology concentrated on apodization techniques for small inner working angle vector vortex coronagraphs[18,19] and the use of (single-mode) fiber-fed high-resolution spectrographs to distinguish planetary signals from starlight residuals[20,21]. The High-contrast imager for Complex Aperture Telescope (HiCAT testbed), located at the Space Telescope Science Institute, specialized in wavefront sensing and control as well as amplitude and phase apodization techniques for high-contrast coronagraphy on segmented or heavily obscured apertures[22-26].

To our knowledge, vacuum coronagraphic results have only been obtained at three facilities. The most prolific one is NASA's High Contrast Imaging Testbed (HCIT) located at the Jet Propulsion Laboratory. It is currently the only vacuum coronagraphic facility that has demonstrated coronagraphic dark holes over spectral bandwidths of 10% or more with contrast levels deeper than $10^{-7}$ at 3–4 $\lambda$/D separations. The HCIT is a set of vacuum coronagraphic testbeds that has been made available for coronagraphic experiments conducted by the US community under the NASA Strategic Astrophysics Technology (SAT) program over the last ~ 15 years. It has been used for testing many different coronagraphic approaches, including some of those mentioned above. The HCIT was also used for key milestone demonstrations in support of the Roman Coronagraph flight instrument development between 2014 and 2020. It currently hosts two "decadal survey testbeds" (DSTs[27,28]) specifically designed to push coronagraphic contrast down to the levels required by HWO. The DSTs are the state-of-the-art in vacuum testbeds based on the accumulated experience of the HCIT facility. The other two facilities are located at the Lockheed Martin Advanced Technology Center, where vacuum coronagraphic demonstrations occurred in 2014–2015 in support of the EXCEDE study[29], and at the University of Arizona, where a new vacuum chamber was recently commissioned, with coronagraphic demonstrations so far limited to a spectral bandwidth of 2% or less (Ewan Douglas, priv. comm.).

Recent laboratory results with starshades have been obtained in air with the Princeton starshade testbed. As with internal coronagraphs, we present here only broadband results, which were obtained over a spectral bandwidth of 12%.

In this paper we collect in one place the recent starlight-suppression results most relevant to HWO, in order to assess (1) the status of the field and (2) the work still needed to reach the performance notionally required by HWO.

In Sec, 2, we review the instrumental contrast performance currently demonstrated in the laboratory[ii], covering various telescope types: on- and off-axis monoliths and on- and off-axis segmented apertures. In line with the high-level science objectives of HWO, and anticipated needs of HWO's starlight suppression system, we only consider approaches that have demonstrated deep contrast in the lab over instantaneous bandwidths of ~10% or more, and at angular separations <= $4\lambda/D$. Polychromatic performance is itself often a good indicator of coronagraph (and testbed) maturity, as achromatizing coronagraphic masks and simultaneously controlling both amplitude and phase across large bandwidths with deformable mirrors remains challenging.

In Sec. 3, we introduce the other two key performance parameters (KPPs) of a starlight suppression system beyond instrument contrast: the off-axis throughput for close-in exoplanet light detection, and the post-calibration contrast obtained after removing residual starlight from science images. Given the extreme levels of starlight suppression required, laboratory experiments have mainly concentrated so far on demonstrating deep contrast at small angular separations. However, off-axis planet signal transmission and post-calibration capabilities become the crucial performance parameters once a threshold raw contrast performance has been reached.

In Sec. 4, we start exploring the relationship between the three starlight suppression system KPPs and HWO's exoplanet science capabilities. As an initial simple step in that exploration, we compare the signals expected from exoearth targets, (exo)-zodiacal dust structures and starlight (attenuated by variable factors) as a function of stellar distance and wavelength. We then adopt as a representative fiducial target the case of an Earth twin seen at quadrature around a Sun twin at 12pc and compute the exposure times required for photometric detection and visible spectroscopy of that target under various KPP levels, including those currently demonstrated in the lab. This provides an initial mapping of the high-level starlight suppression KPPs values – or combinations thereof – required for viable scientific observations and helps identify possible trades between individual KPPs. For instance, it provides some insight on the relative science value of improving raw contrast vs increasing off-axis throughput or vs improving raw speckle calibration.

The actual system implementation trades, including detailed telescope design and coronagraph masks selection, as well as higher system level trades of, e.g., telescope wavefront stability (or knowledge) vs. coronagraph resilience to aberrations are beyond the scope of this paper. The contrast sensitivity of different coronagraphs to telescope pointing jitter (and hence stellar diameter), aberrations changes[iii] caused by vibrations or slow thermal drifts is a crucial characteristic to include in future implementation trades. It is being investigated separately[30] and will be discussed in a separate publication (Por et al. in preparation).

We end in Sec. 5 with a list of suggested key improvements, high priority technical trades and immediate laboratory demonstrations needed to develop starlight suppression systems towards the notional exoplanet science needs of the future HWO mission.

---

[ii] Also see the regularly-updated compilation of testbed performance demonstrations maintained by Brendan Crill on behalf of the Exoplanet Exploration program located at: https://exoplanets.nasa.gov/internal_resources/2595

[iii] At the $10^{-10}$ level, the contrast performance degradation caused by even tens of picometers of uncorrected wavefront changes will generally be noticeable by coronagraphs, and it will differ among them. The effect will be much less on starshades, which can tolerate nanometers of drifts before seeing any contrast degradation.

## 2    Starlight Suppression Laboratory Experiments and Raw Contrast Results

Once the exoplanet science mission objectives are clearly identified, corresponding yield metrics and quantitative figures of merit can be defined, e.g., the number of exoearths spectra obtained over a specified wavelength range at a desired signal-to-noise ratio (SNR) and spectral resolution within some total observing time, and a close-to-optimum mission operations concept can be derived to maximize science yield[31-34]. Assuming that the telescope diameter and list of nearby Sunlike target stars to draw from are fixed, the exoearth science yield is essentially set by three starlight suppression system characteristics (KPPs), which vary as a function of spectral bandwidth and location around the star[35]:

- "Raw contrast", defined as the instrumental contrast measured in the dark hole region of the science images *before* any differential imaging and other post-calibration techniques are applied. The raw contrast is field dependent and, for a given stellar diameter,  is a property solely of the optical system. At any given position in the dark hole, a planet with a planet-to-star flux ratio equal to the local raw contrast value would off produce the same number of counts as observed at that location from the partially suppressed star. To minimize the exposure time required for exoearth observations - especially spectroscopy -, the raw contrast value must be low enough that the irreducible photon noise associated to other astrophysical background sources, such as zodi or exozodi dust, dominates over stellar shot noise. In practice, that threshold raw contrast level varies with stellar distance, wavelength and exozodi brightness, and ranges from $\sim 10^{-10}$ to $\sim 10^{-8}$ (Sec. 4.1).
- Post-calibration contrast, also field-dependent, defined as the minimum planet-to-star flux ratio detectable with an SNR of 1 *after* calibration and processing of the raw science images (Sec. 3.2 and 4.2). E.g., for visible observations of planets with a $10^{-10}$ flux ratio (same as an Earth twin seen at quadrature around a Sun twin) at an SNR of 20, the post-calibration contrast rms error must be $< 5 \times 10^{-12}$.
- Off-axis ("core") throughput, defined as the fraction of off-axis (planet) light transmitted through the starlight suppression system into a specified off-axis PSF core area (Sec. 3.1).  The off-axis core-throughput 2-dimensional map hence captures both the effects of PSF sharpness and finite transmission close to the star. This metric is more informative than the starlight suppression system inner working angle (IWA), which is defined as the angular separation at which the system off-axis transmission reaches 50% of its maximum off-axis value. While the IWA is a useful performance parameter, it can also be deceptive when used alone. For example, a system with a larger (= worse) IWA but a higher maximum off-axis transmission may still outperform a system with a smaller (= better) IWA but a lower off-axis transmission.

All lab demonstrations discussed hereafter have been carried out by investigators selected under the auspices of the NASA SAT program and were conducted at visible wavelengths. Except for the HiCAT results, which were obtained in air, all other coronagraphic performance results reported in this section were obtained in vacuum in the HCIT, with passive mechanical vibration isolation and active correction of residual wavefront aberrations, providing a space-like environment that is turbulence-free and (mostly) thermally stable. While there are still some small dynamical effects, e.g., fast residual pointing jitter at the milliarcsecond level, the vacuum lab performance results reported hereafter can then essentially be seen as "static" results. Finally, for all coronagraphic results presented hereafter, two types of deformable mirrors (DMs) were used: 48x48 actuators DMs from Adaptive Optics Associates (Xinetics AOX), or ~1000 to ~2000 actuator DMs from the Boston Micromachines Corporation (BMC).

## 2.1 Starlight Suppression System Contrast Definitions

Before describing the results achieved in the laboratory, it is worth noting that the contrast level of a starlight suppression system is generally expressed in terms of either *normalized intensity* or *raw contrast*, which correspond to two different quantities.

For any pixel in the dark hole, the *normalized intensity* is defined as the ratio of the number of residual starlight counts measured in that pixel when the star is on-axis, with all starlight suppression optics in the beam path, to the maximum number of counts per pixel measured under the exact same conditions, but with the coronagraphic focal plane mask (or the starshade mask) removed to let starlight through. Note that the denominator in this ratio is the same for all pixels, meaning that the normalized intensity map is a scaled version of the observed 2D coronagraphic image. Normalized intensity is the quantity of choice when assessing the degree of starlight suppression. But it says nothing about the system's ability to transmit the signal of off-axis sources (e.g., exoplanets or extended disk structures). All testbed contrast results presented in this section are expressed in terms of normalized intensity, unless specified otherwise.

On the other hand, the "*raw contrast*" (RC) computed at any pixel in the coronagraphic dark hole accounts for spatial variations in the off-axis transmission of the focal plane mask. It is defined as the ratio of the number of residual starlight counts measured in a pixel (or photometric aperture centered around it) when the star is on-axis with all starlight suppression optics in, to the number of counts measured in that same pixel (or photometric aperture) under the same conditions, but after shifting the input star to that pixel. Physically, in a given pixel, the raw contrast value corresponds to the planet flux relative to the star that would produce the same signal through the system as the residual starlight. The raw contrast is then a more relevant quantity than normalized intensity when assessing the detectability of exoplanets or extended circumstellar disks.

The raw contrast per pixel is equal to the normalized intensity per pixel divided by the focal plane mask (2D) off-axis transmission response. This means that raw contrast is always worse than normalized intensity, and that it becomes infinite on-axis. The two figures of merit are only equal (approximately) for sources far enough off-axis to be unaffected by the focal plane mask, yet close enough to the axis to be unaffected by the edges of any downstream field stop. A comparison of normalized intensity and raw contrast values as a function of angular separation is given in[35] for various coronagraphs. For the starlight-suppression systems considered here, the normalized intensity and raw contrast values only differ by a few tens of percent or less for separations greater than $3\lambda/D$. We also note that the raw contrast in a multi-pixel photometric aperture is generally worse than the raw contrast per pixel, because the off-axis PSF maximum pixel level is higher than the mean off-axis PSF level in the photometric aperture.

## 2.2 Coronagraph Demonstrations for Off-axis Monolithic Telescopes

In terms of fundamental physics limits, coronagraph raw contrast performance is not strongly dependent on aperture shape, obstructions, or segmentations[36]. However, all else being equal, a plain circular aperture without any obscuration, as given by an off-axis telescope, is more straightforward to work with than one with obscurations, and the best broadband coronagraphic laboratory results to date have been achieved for this type of aperture. They were obtained at JPL's HCIT, and fairly similar levels of raw contrast performance were reached with a "classical" Lyot coronagraph, a hybrid Lyot coronagraph and a vector vortex coronagraph. For completeness, we also present the broadband results obtained with a phase-induced amplitude apodization coronagraph, which demonstrated worse contrast but accessed smaller angular separations.

### 2.2.1 Classical Lyot coronagraph (CLC)

The conceptually simplest coronagraph design is the classical Lyot coronagraph (Fig.1) invented by Bernard Lyot for observations of the solar corona nearly a century ago. On-axis starlight cancellation is achieved via the combination of an opaque circular mask located in an intermediate focal plane and an aperture stop located in a downstream pupil plane conjugate. The downstream aperture stop, called a Lyot stop, is used to reduce the amount of starlight diffracted inside the pupil by the sharp edges of the focal plane mask. As the focal plane mask (FPM) gets smaller, a larger fraction of the aperture needs to be blocked by the Lyot stop to reduce the FPM diffractive effects. This means that for a CLC, a compromise must be made between the ability to detect sources at small separations and coronagraph throughput. Modern coronagraphs, such as the ones described in the next sections, use pupil apodization or phase mask techniques to overcome that limitation and theoretically provide higher sensitivity at smaller separations than a CLC.

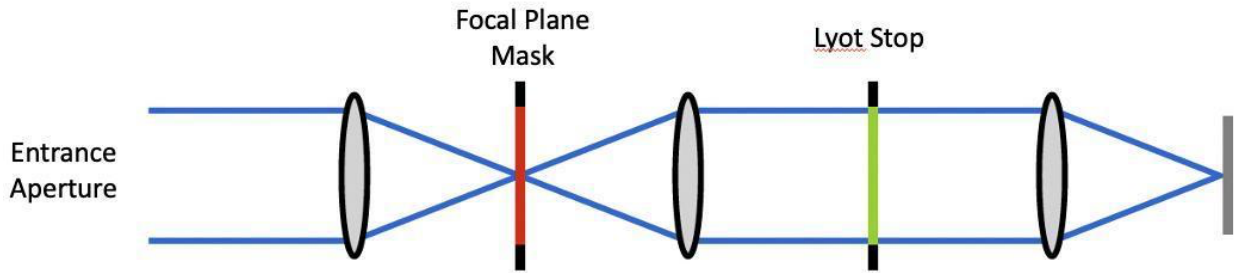

**Fig. 1. Schematic optical layout of a "classical" Lyot coronagraph, with a hard-edge focal plane mask (FPM) inserted in an intermediary focal plane and a Lyot stop located in a pupil plane conjugate. For the DST CLC experiment considered here[37], the entrance pupil consisted of an unobscured circular aperture and the FPM size was set to 2.7 λ/D in radius. In order to control starlight diffraction over the 10% spectral bandwidth, the Lyot stop blocked any light located outside of a 0.2D – 0.675D annulus, resulting in significant transmission losses.**

However, given its relative simplicity to set up and model, a CLC was tested with an unobscured, circular pupil using the Decadal Survey Testbed (DST). The DST has two 48x48 DMs from Adaptive Optics Associates (Xinetics = AOX) and is equipped with a low-order wavefront sensing and control (LOWFS/C) subsystem to sense and correct residual dynamic wavefront disturbances in the vacuum chamber (see Section 3). Conversely to the hybrid Lyot mask coronagraph presented in the next subsection, the focal plane mask is a simple binary mask in amplitude, with no spatially varying phase effects purposely introduced by the mask itself. For this reason, we refer to it as a "classical" Lyot coronagraph, although the two DMs are used to achieve a complex-valued apodization of the pupil.

For this demonstration, the Lyot stop was quite lossy: it blocked any light coming from outside of 67.5% of the original beam radius, and from inside the inner 20%. Blocking both the inner and outer regions of the beam was required to achieve broadband wavefront control with the DMs and deep broadband starlight cancellation (G. Ruane, private communication). Using this set-up, an average starlight suppression level of ~3.8 x 10⁻¹⁰ (normalized intensity) was demonstrated over a 360 degree dark hole region covering 3–8 λ/D in separation, using linearly polarized broadband light (10% bandwidth centered at 0.55 μm) and a linear polarization analyzer[37]. The resulting high contrast image is shown in Fig. 2 (left panel). Further tests and analysis suggest that ~80% of the observed contrast is limited in nearly equal parts by four different effects: DM actuator finite stroke resolution (due to 16 bits electronics at the time); chromatic control residual; focal plane occulter ghost; and residual line-of-sight jitter. The origin of the last 20% contribution (about 5 x 10⁻¹¹) to the observed contrast floor is unknown, implying that other processes must be considered.

A more recent, higher throughput, CLC experiment achieved similar contrast performance over a 20% spectral bandwidth with a similar DST laboratory set-up, but over a one-sided[iv] dark hole (Fig. 2, right panel) extending from 5 to 13.5 λ/D, i.e., located at larger separations[38].

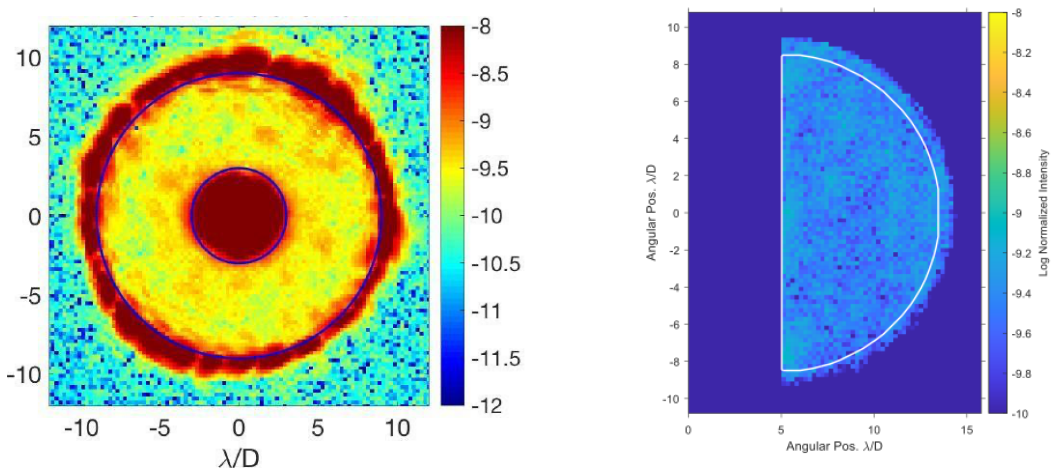

**Fig. 2. Left (adapted from[37]):** CLC results obtained in the HCIT DST testbed using linearly polarized broadband light (10% bandwidth centered at 0.55 μm). The average contrast (normalized intensity) measured over the 360 degree dark hole for separations ranging from 3 to 8 λ/D was 3.82 x 10⁻¹⁰. **Right (adapted from[38]):** DST CLC results obtained with the same set-up but over a 20% bandwidth centered at 0.56 μm on a one sided dark-hole extending from 5 to 13.5 λ/D. The average contrast measured over that one-sided dark hole was 3.97 x 10⁻¹⁰.

---

[iv] As the spectral bandwidth increases, coronagraphic starlight cancellation requires more degrees of freedom (i.e, DM actuators) to control phase and amplitude, or may only be achieved over a smaller dark-hole region. Different dark-hole extents may be considered for initial exoplanet blind search detections (full dark hole) vs follow-up spectroscopic observations (e.g., half dark hole).

*2.2.2 Hybrid Lyot coronagraph (HLC)*

In order to improve the coronagraph off-axis throughput at small separations, the FPM spatial transmission profile should be apodized, rather than constant in the CLC case. For a hybrid Lyot coronagraph (HLC), the optical layout is the same as in Fig. 1. But the FPM amplitude transmittance is now varying spatially, as well as its induced phase shift. The FPM amplitude and phase profiles are jointly optimized with the DM solution to minimize contrast over a specified dark hole region and spectral range.

The best HLC 10% bandwidth laboratory results reported to date (Fig. 3, right panel) used a linear occulting mask[39,40]. The demonstrated contrast was 6.0 x 10^-10 in the inner 3-4 λ/D field, and the spatially averaged contrast was 5.2 x 10^-10 in the outer 3-15 λ/D field. The FPM complex apodization was provided by thickness-profiled metallic and dielectric films superimposed on a glass substrate. These 10% contrast results are fairly similar to those achieved with the CLC, but obtained under different operating conditions (one-sided vs full dark hole, dual polarization vs single polarization, and single DM vs two DMs), making an absolute comparison difficult.

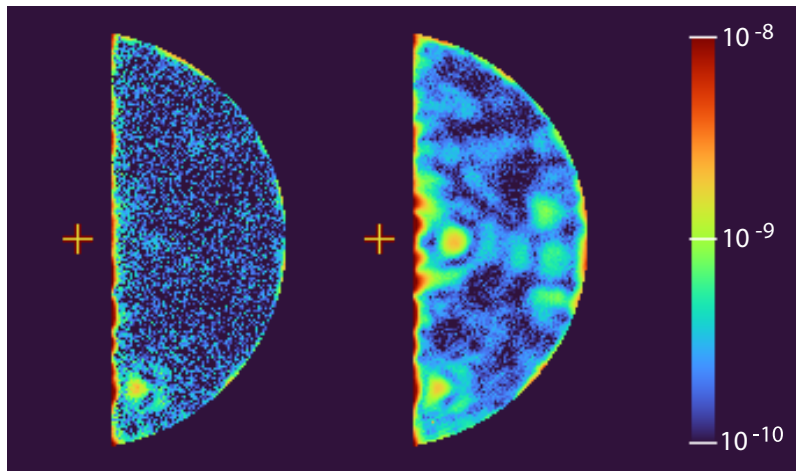

**Fig. 3. HLC results obtained in the HCIT testbed using unpolarized light (left: 2% bandwidth, right: 10% bandwidth) centered at 0.80 μm. The average contrast (normalized intensity) measured over the one-sided dark hole, which extends from 3 to 15 λ/D, was 2.0 x 10^-10 in 2% bandwidth and 5.2 x 10^-10 in 10% bandwidth. The average contrast measured in the inner 3 to 4 λ/D region was 3.2 x 10^-10 in 2% bandwidth and 6.0 x 10^-10 in 10% bandwidth.**

Compared to the 10% spectral bandwidth case, contrast results obtained with the same HLC mask were ~2x better over 2% bandwidth and ~3x worse with 20% bandwidth. The observed degradation of contrast performance with increased bandwidth, together with a model analysis of the observed speckles morphology and inspection of fabrication hardware all pointed to a calibration error for the FPM dielectric thickness, an issue that has since been remedied.

*2.2.3 Vector vortex charge 4 coronagraph (VVC4)*

There are two kinds of optical vortex known: the scalar optical vortex[41,42], implemented by a structural (i.e., variable thickness) helix providing a scalar optical phase delay which applies to the two orthogonal polarization components of natural light identically, and the vector vortex coronagraph[43-46], implemented by a rotationally symmetric half-wave plate (HWP), providing a "geometrical" phase shift that applies opposite phase screws to the two orthogonal circular polarization states.

By contrast to the Lyot coronagraphs, the vortex focal plane mask is a transparent optic, which imparts a spiral phase shift of the form exp($il\varphi$) on the incident field, where $l$ is an even nonzero integer known as the "charge" and $\varphi$ is the azimuth angle in the focal plane. Light from an on-axis source (i.e., the star) that passes through a *circular unobscured* entrance pupil of radius $a$ and the transparent vortex focal plane mask is completely diffracted outside of the downstream Lyot stop of radius $b$, with $b<a$, providing a high-throughput solution with perfect starlight suppression in the ideal case. Fig. 4 shows the schematic of a vortex coronagraph layout, which is seen to be identical to the CLC/HLC layout except for its phase-based focal plane mask.

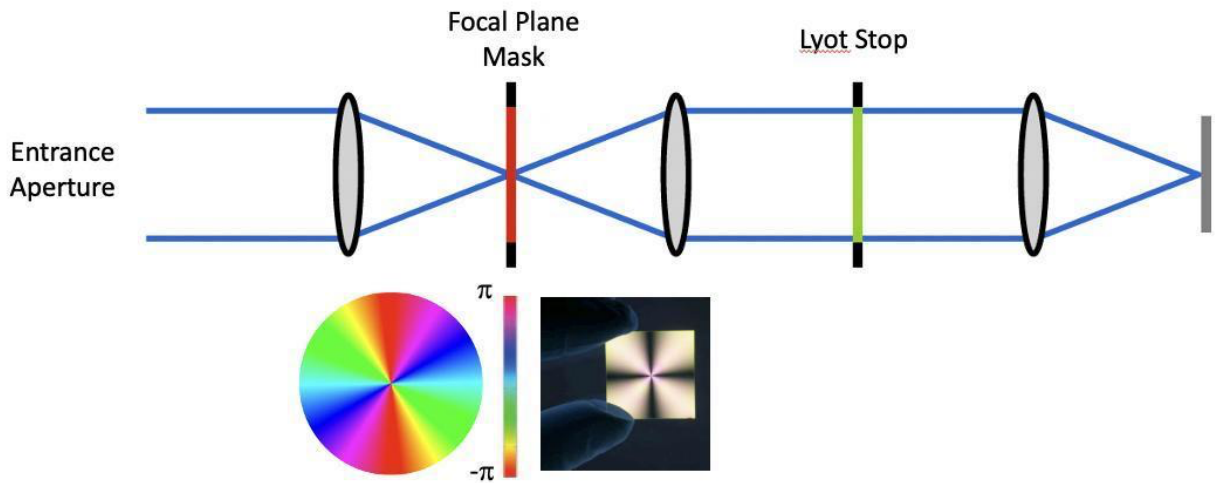

**Fig. 4. Schematic layout of a vortex coronagraph, including its focal plane phase mask of complex transmittance exp($il\varphi$) (illustrated with $l$=2) and circular Lyot stop. For the results described in this section, the entrance pupil consisted of an unobscured circular aperture. A vector vortex of charge $l$=4 was used, meaning that the focal plane phase rotated twice faster around the axis than represented in the figure, and the Lyot stop opening diameter was set to 80% of the entrance pupil diameter.**

In addition, because the focal plane phase mask is transparent instead of opaque, the vortex coronagraph provides high throughput for off-axis point-like sources, i.e., planets located at small angular separations from the star. In practice, as its charge increases, an optical vortex coronagraph becomes less sensitive to low-order aberrations[46], but its throughput at small angular separations also decreases. As shown by previous studies[2,3], vortex charges of four or six provide a reasonable compromise between the desired resilience to low-order aberrations and high sensitivity to close-in, e.g., habitable zone, exoplanets. A charge six VVC was ultimately preferred in the HabEx case, as it exhibited far less sensitivity to tip-tilt and astigmatism, the main low-order aberrations induced by polarization cross-talk effects in the HabEx telescope beam train[47]. It is also worth noting that while the theoretical performance of vortex coronagraphs does not fundamentally change for segmented primary mirrors, assuming the segment co-phasing and pupil apodization requirements can be met, their throughput and robustness to wavefront errors is currently found to degrade significantly on centrally obscured (on-axis) telescopes unless hybridized with the PIAA coronagraph[48].

Deepest rejection has been obtained with an unobscured circular aperture with a vectorial vortex coronagraph of charge four (VVC4) in the DST using the same two 48x48 AOX DMs as in the CLC case. In this case, the vortex geometrical phase ramp was implemented using birefringent liquid crystal polymer waveplates, with their fast axis orientation rotating as a function of azimuth angle. The contrast (normalized intensity) achieved by the system on linearly polarized light was $1.6\times10^{-9}$ and $5.9\times10^{-9}$ in 10% (Fig. 5) and 20% optical bandwidths, respectively, averaged over 3 to 10 $\lambda$/D separations on *one side* of the pseudo-star[49]. The residual starlight intensity was shown to be coherent (i.e., correlating with changes in DM surface) and likely dominated by spatial imperfections in the vortex focal plane mask. As a result of these imperfections, the normalized intensity scales roughly with the square of the spectral bandwidth for bandwidths > 10%. Future work aims to achieve $5\times10^{-10}$ contrast in a 10% bandwidth using a new generation of vector vortex masks with tighter tolerances on their imperfections and defects, improved suppression of the polarization leakage and a new set of higher-resolution DM controller electronics to reduce quantization errors.

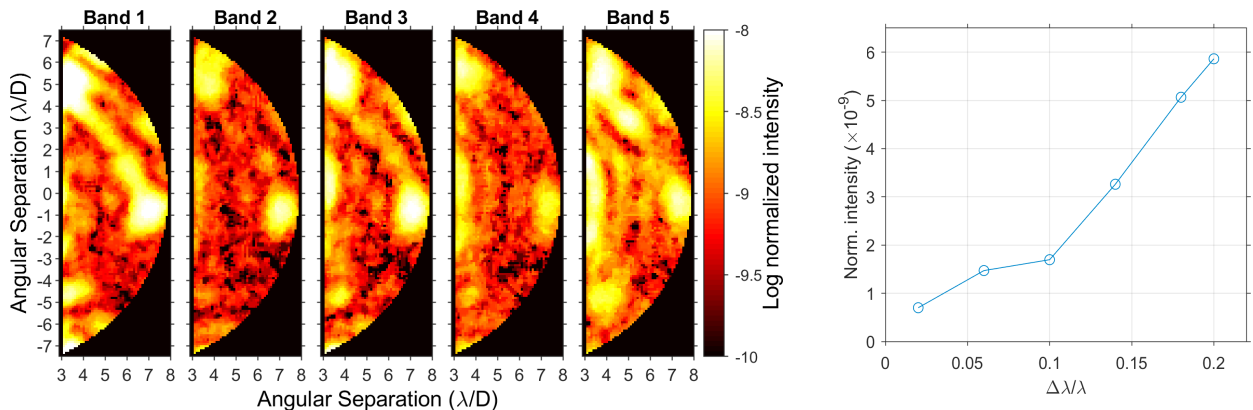

**Fig. 5. Left: Raw normalized intensity images obtained in five 2% sub-bands with a VVC4 operating on an unobscured circular aperture. All five images are obtained with the same DM settings, determined to minimize the total starlight residuals over the synthetic 10% bandwidth centered around 0.65 μm. The resulting normalized intensity spatial average over the *one-sided* 3 to 10 λ/D dark hole is 1.6 x 10⁻⁹. Right: Spatial average of normalized intensity measured over the dark hole with the same VVC4 set-up but optimizing the DM settings for spectral bandwidths ranging from 2% to 20%. Adapted from[49].**

In addition to this dual DM VVC4 experiment, tests were also conducted with a single BMC DM, also using polarized light over a 10% bandwidth[53]. The raw contrast degraded by a factor of 4 to 5 in the inner 3-5 $\lambda$/D region, and by a factor of ~2 over the full one-sided 3-10 $\lambda$/D dark hole, with an average contrast measured at 3.2 x 10⁻⁹. While the VVC4 clear aperture results are worse with a single DM than two, they provide the relevant point of comparison for the VVC4 segmented mask results presented in Sec. 2.3, which were obtained with the same single DM.

*2.2.4. Phase-induced amplitude apodization (PIAA) coronagraph*

Phase-Induced Amplitude Apodization[50] (PIAA) is an alternative to classical pupil apodization techniques that use a lossy amplitude pupil mask. In the case of PIAA, an achromatic apodized pupil can be obtained by reflection of an unapodized wavefront on two aspheric mirrors (the "PIAA optics") carefully shaped to redistribute the light in the pupil. The surface quality of the PIAA optics and their alignment are critical for optimum performance. Provided those can be achieved, the PIAA approach theoretically provides a small inner working angle than the previous coronagraph types and - because the phase-induced apodization is nearly lossless - a high coronagraphic throughput with almost no loss in PSF sharpness close to the optical axis. It is worth noting that because of the intervening PIAA optics (with only the first PIAA mirror located in a pupil plane), the PSF is no longer translation invariant and becomes less concentrated for off-axis sources. That image distortion becomes significant for separations larger than a few $\lambda/D$ but can be remedied by adding a set of "inverse PIAA optics" downstream of the coronagraph masks, essential to reform the telescope entrance pupil. Because starlight is already canceled at that point, the inverse PIAA optics surface requirements are not as stringent as for the upstream set of PIAA optics. But they still add two extra reflections and some complexity to the system (Fig. 6).

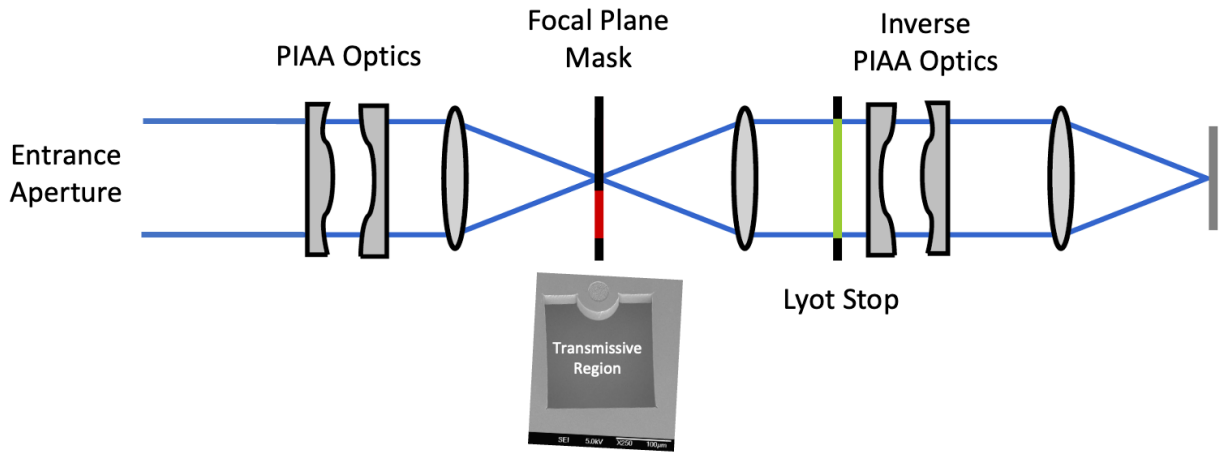

**Fig. 6. Schematic layout of a PIAA coronagraph, including the two PIAA optics (aspheric mirrors) generating the beam apodization, and the inverse set of PIAA optics used to reform the entrance pupil downstream of the coronagraph to correct for off-axis PSF distortion. For the HCIT experiment, the incoming beam into the PIAA optics was diverging, and the focal plane mask limited the dark hole extent to 2 to 4 $\lambda/D$ (transmissive region of inserted SEM image) to concentrate on the performance achieved at small separations. No inverse PIAA optics were used or necessary to correct distortions that close to the optical axis.**

A PIAA coronagraph was tested in vacuum in the HCIT in 2013 on a circular monolithic aperture using a single DM for wavefront control[51,52]. The testbed reached $10^{-8}$ mean broadband (10% bandwidth) contrast averaged over a one-sided dark hole extending from 2 to 4 $\lambda/D$ (Fig. 7) with a linear polarizer inserted immediately before the science camera. Interestingly, the monochromatic contrast demonstrated at 0.8 µm with the same set-up was 5 x $10^{-10}$, and the significant degradation in performance with spectral bandwidth is believed to be due to surface errors on the PIAA mirrors.

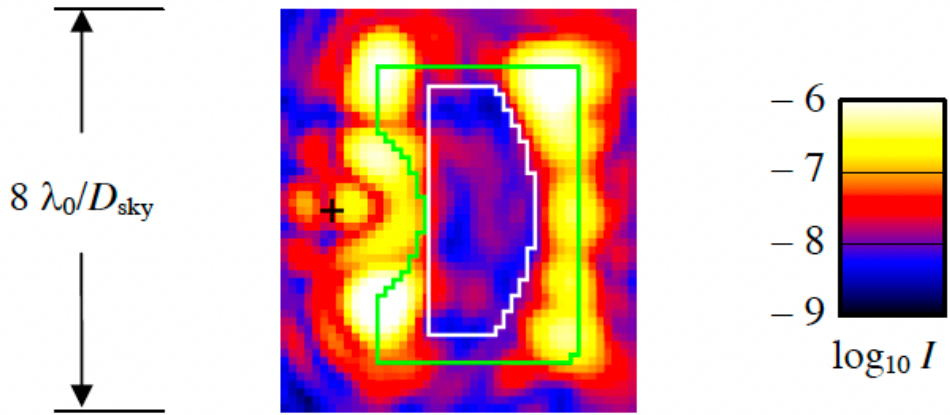

**Fig. 7. PIAA results obtained in the HCIT over a 10% spectral bandwidth centered around $\lambda_0 = 0.8$ μm. The contrast image extends from −1 to +6 $\lambda_0/D$ in x, +/− 4 $\lambda_0/D$ in y. The source center ("star" position) is marked with a small black cross, and the "scored" region border with inner edge at x = 2 $\lambda_0/D$, outer radius 4 $\lambda_0/D$ is shown in white. The green line marks the edge of the occulter, the boundary between being fully opaque and fully transmitting. The mean contrast measured over the dark hole region is $10^{-8}$, to be compared to $5 \times 10^{-10}$ in monochromatic light.**

## 2.3 Coronagraph Demonstrations for Off-axis Segmented Telescopes

After off-axis monoliths, the next most favorable aperture for coronagraphy, from the standpoint of current coronagraph designs and testbed results, is a segmented off-axis aperture. As in the monolithic off-axis case, this entrance pupil has no central obscuration, but it is discontinuous and slightly obscured by segment gaps. The best broadband vacuum results to date with this aperture type have been obtained with a vector vortex coronagraph (VVC4) using a segmented mask to simulate the amplitude effect of segment gaps on the entrance aperture (Section 2.3.1). No HLC tests have been conducted to date in vacuum on segmented apertures. The only segmented aperture Lyot results have instead been obtained in air, with a phase-apodized-pupil Lyot coronagraph (PAPLC) (see section 2.3.2 for definitions). The PAPLC results used a segmented deformable mirror as an actual discontinuous wavefront, allowing simulation of the effects of both amplitude and phase discontinuities between telescope segments. The VVC4 and PAPLC results are described in the following two sections.

### 2.3.1 Vector vortex charge 4 coronagraph (VVC4)

The above-described VVC4 was also tested in the HCIT DST[53] with a hexagonally segmented pupil mask located at the entrance of the coronagraph (Fig. 8, left panel), representing a - perfectly phased - off-axis segmented entrance pupil with five segments across. The width of the segment "amplitude gaps" was about 60 microns or 0.2 % of the diameter. In this case, a single BMC DM with ~2000 actuators was used for wavefront correction, and the Lyot stop inner opening diameter was increased to 0.87 D.

The effect of segmentation was initially tested at a single wavelength using a 0.637 μm diode laser. Interestingly, the monochromatic contrast performance was found to be the same with or without segmentation of the entrance aperture, and this is the reason we mention it here. The monochromatic contrast (normalized intensity) demonstrated with this set-up was 3.6 x 10⁻¹⁰ averaged over a 3 to 10 λ/D one-sided dark hole. Starlight residuals were dominated by slight Airy rings centered on the star (Fig. 8, middle panel) that did not react to DM actuation. They were attributed to polarization leakage caused by imperfect retardance of the vector vortex mask, or the finite extinction ratio of the circular polarizer-analyzer pair used to isolate a single circular polarization through the system.

Using the segmented pupil in broadband light (10% bandwidth centered at 0.650 μm), the spatially-averaged normalized intensity over the dark hole increased about tenfold, to 4.7 x 10⁻⁹. The observed degradation with bandwidth is thought to be primarily caused by chromatic errors in the vector vortex mask itself[53]. However, the broadband performance observed in the segmented case is also 1.5 x 10⁻⁹ worse than the 3.2 x 10⁻⁹ achieved with the exact same single-DM broadband VVC4 set-up in the monolithic pupil case. So, while the demonstrated VVC4 contrast performance always degrades as the spectral bandwidth increases, that degradation appears slightly more pronounced in the segmented case than in the monolithic one. More tests will be required to disentangle the impact of mask defects from the intrinsic effects of segmentation on broadband VVC performance.

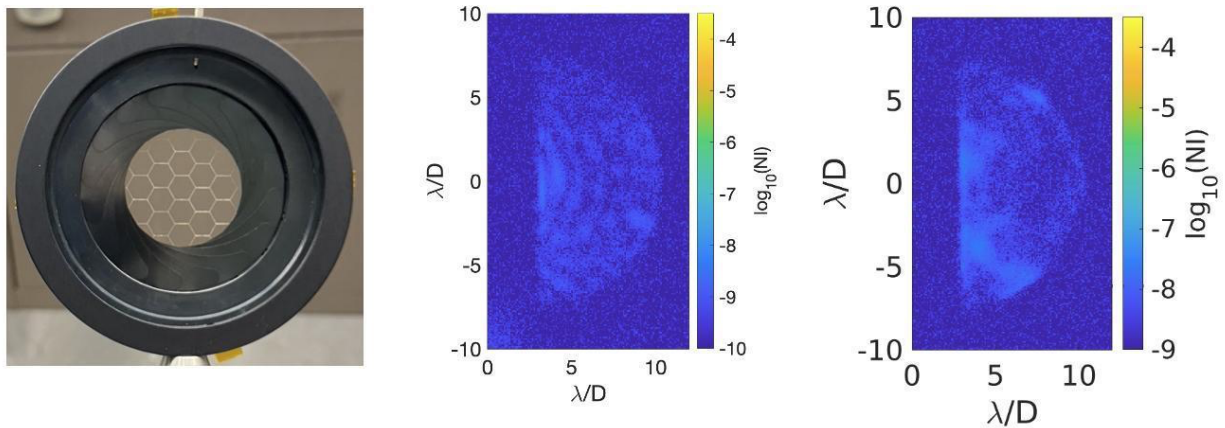

**Figure 8. Left: Entrance aperture used in the segmented VVC4 HCIT/DST lab experiments. Middle: Raw normalized intensity images obtained using a 0.637 μm laser diode. The normalized intensity spatial average measured over the *one-sided* 3 to 10 λ/D dark hole is 3.6 x 10⁻¹⁰. Right: Raw normalized intensity images obtained over a 10% spectral bandwidth centered at 0.650 μm. The normalized-intensity spatial average measured over the *one-sided* 3 to 10 λ/D dark hole is 4.7 x 10⁻⁹. Adapted from[53].**

For this HCIT VVC4 demonstration, the diffractive effects of segment gaps were suppressed using phase apodization provided by the DM. An alternative approach used an amplitude-apodized vector vortex (AVC), tested in air in Caltech's HCST. In this case, an efficient apodizer was manufactured using carbon nanotubes acting as microdots to encode a grayscale pattern. The in-air AVC set-up demonstrated a spatial mean contrast of 8.4 x 10⁻⁹ over a one-sided dark hole extending from 5 to 10 λ/D using polarized light over a 10% spectral bandwidth[18]. There are on-going developments to repeat such AVC demonstrations in vacuum and extend them to smaller separations consistent with HWO needs.

### 2.3.2. Phase-apodized-pupil Lyot coronagraph (PAPLC)

The Phase-Apodized-Pupil Lyot Coronagraph (PAPLC) inherently produces a one-sided dark hole[54]. Like other one-sided dark holes presented above, it is better suited to exoplanet spectral characterization than initial blind searches and broadband detection of exoplanets. It uses pupil-plane phase apodization to apodize the PSF falling on an (off-set) knife edge focal plane mask to create a one-sided dark zone with inner working angles theoretically as small as 1.5 λ/D. The apodizer can be implemented using a phase-only mask or using a pair of deformable mirrors. The PSF is not aligned with the edges of the focal plane mask, but rather offset by 1 to 2 λ/D. With the phase-only apodizer, this offset is applied using the apodizer phase map itself, producing a PSF whose position is invariant with respect to the knife edge, and creating an inherently achromatic coronagraph. With other apodizers, the offset needs to be applied by moving the PSF on the focal plane mask. Naïve designs already provide robustness against aberrations along the knife edge direction. With more advanced designs of the apodizer, this can theoretically be extended to all low-order aberrations up to the specified order.

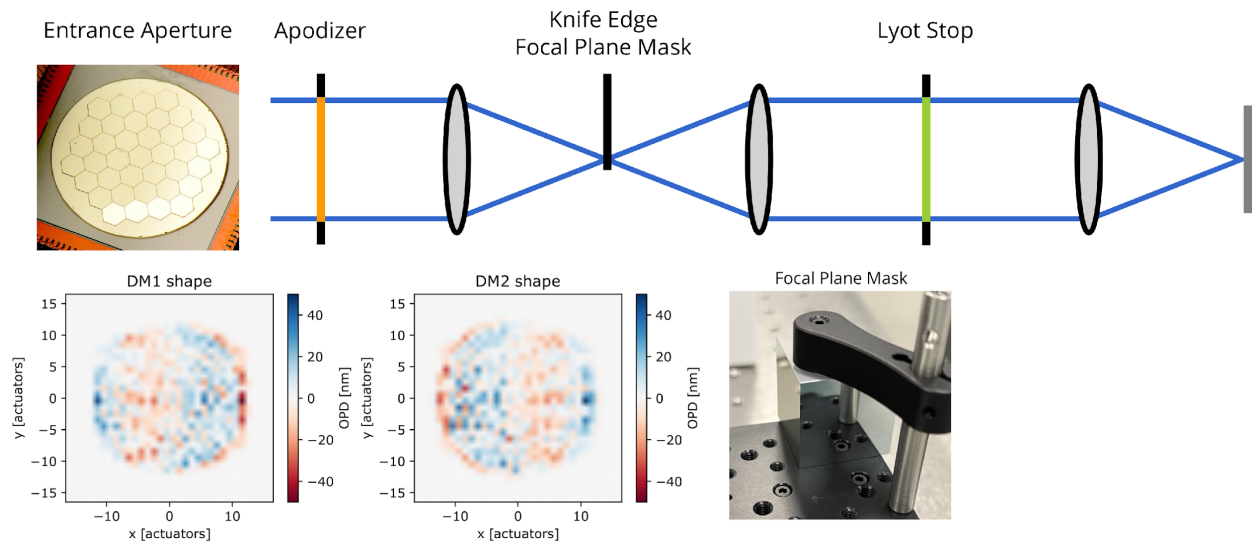

**Fig. 9. Conceptual layout of the PAPLC design tested in air on the HiCAT testbed. The entrance aperture, produced by the IrisAO deformable mirror, contains 37 individually movable hexagonal segments. Two Boston Micromachines Corporation Kilo-DMs are used to create the phase-induced apodization (DM maps shown) and to perform high-order wavefront control. Note that instead of reflection off the phase knife, this figure schematically shows the beam propagating to the Lyot stop in transmission.**

The PAPLC laboratory tests were conducted in air, at STScI's HiCAT testbed. The HiCAT light source illuminates the pupil mask (Lenox Laser), the IrisAO segmented deformable mirror, the reflective apodizer (a fold mirror for the PAPLC), and two continuous Boston Micromachine Corporation Kilo-DM deformable mirrors, each with 952 actuators. These deformable mirrors are used for both the apodizer of the PAPLC and for high-order wavefront control. The light is then focused onto the edge of a right-angle prism (Thorlabs) forming the focal plane mask, of which the reflected part is collimated onto a circular Lyot stop (Lenox Laser). Finally, the light is focused onto the science camera. HiCAT has not seen polarization differences at the contrast levels accessible in air so far and operates in unpolarized light. A 37-segment unobstructed hexagonally-segmented input aperture was used for these tests (Fig. 9).

The mean contrast observed in 9% broadband light (centered at 0.66 um) was 4.2 x 10<sup>-8</sup>, from 2 to 13 λ/D, and for 25% broadband light, 9.5 x 10<sup>-8</sup> from 2 to 12 λ/D (Por et al. in prep, Fig. 10). The current performance appears to be limited by chromaticity effects of the apodizer, which should be correctable according to wavefront control simulations.

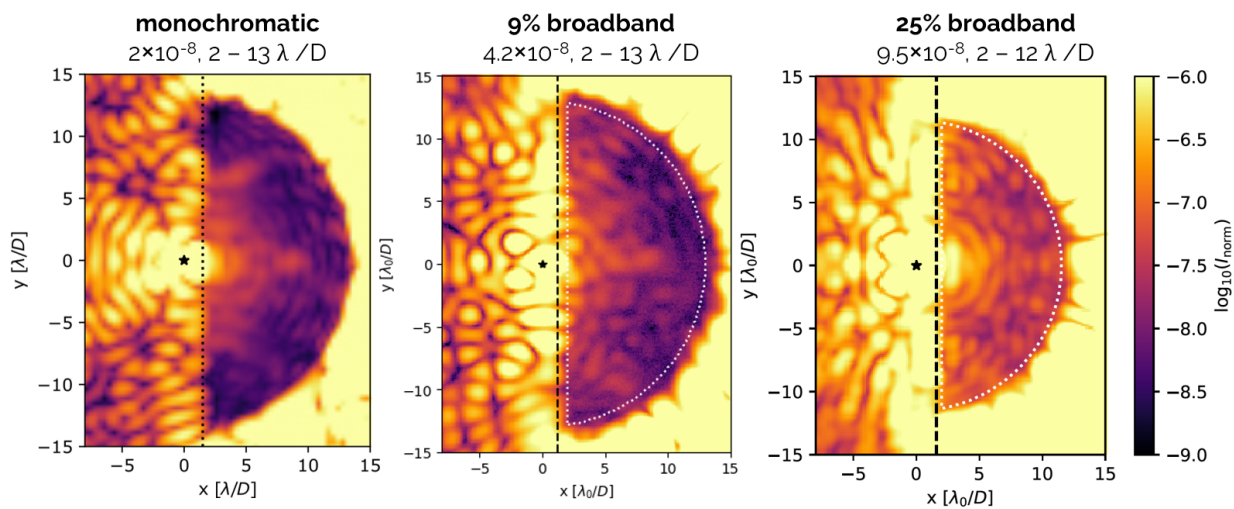

**Fig. 10. PAPLC in-air laboratory results (Por et al. in prep). The vertical dashed line indicates the knife edge location at $2\lambda_0/D$. Middle: 9% bandwidth 1-sided dark hole obtained around $\lambda_0$=0.66 μm. Right: 25% bandwidth 1-sided dark hole obtained around $\lambda_0$=0.66 μm. The mean contrasts observed are given in the text.**

### 2.4 Coronagraph Demonstrations for On-axis Monolithic Telescopes

Broadband vacuum coronagraphic demonstrations for on-axis monolithic apertures have been obtained in the case of the Roman D=2.4m telescope which features a heavily obscured aperture with a 30% central obscuration and six secondary struts that are 0.03D wide. Roman's complex entrance pupil provides a challenge for coronagraph architects and is likely not representative of the HWO aperture. However, we describe it here to illustrate the trade between coronagraphic raw contrast and off-axis core throughput, two of the main starlight suppression parameters defined above. HLC and shaped pupil coronagraph (SPC) masks were developed and optimized for operation with Roman at different visible wavelengths and over various dark hole extents[55]. Both types of masks were tested in the HCIT on a Roman-like pupil using a pair of AOX 48x48 DMs for wavefront control, and demonstrated deep broadband contrast through early laboratory testing on representative Roman technology testbeds[56,57], as summarized hereafter. However, high contrast performance and reasonable resilience to wavefront changes or misalignments were only achieved at the expense of reducing the off-axis core (planet) throughput[58].

We note for completion that a back-up PIAA-based architecture (called PIAACMC, see section 2.5) was also developed for the Roman aperture with the goal of providing high contrast down to separations of only 1.3 λ/D. However, due to limited resources and schedule, and the more complex nature of PIAACMC, it did not undergo sufficient lab testing to determine its ultimate contrast performance[59], and only the HLC and SPC modes will fly with Roman.

### 2.4.1 Roman HLC coronagraph

The layout of the Roman HLC coronagraph is shown in Fig. 11. As in the off-axis monolith HLC case (Sec. 2.2.2), the FPM amplitude and phase profiles are jointly optimized with the DM solution to minimize contrast over a specified dark hole region and spectral range. To reach deep broadband suppression of the starlight diffracted from the heavily obscured Roman telescope pupil, significant wavefront distortion is injected by the DMs, which degrades the off-axis PSF by scattering light from the core into extended wings. Additionally, a significantly undersized Lyot stop is used to control core diffraction. Both effects contribute to a low core throughput, the price to pay to reach and maintain high contrast with the Roman entrance aperture.

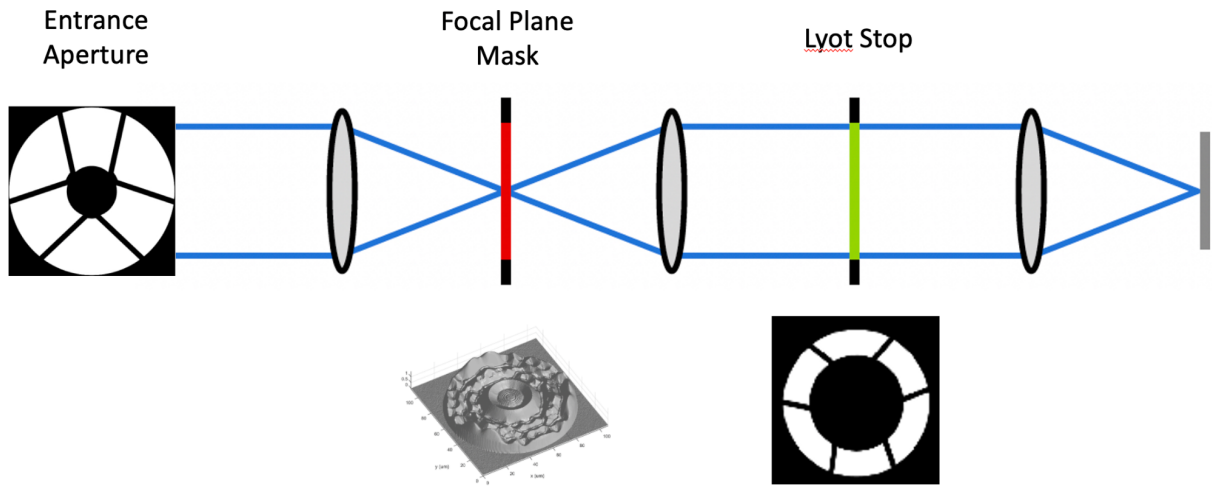

**Fig. 11. Optical layout of the Roman HLC coronagraph, showing the Roman telescope entrance pupil, a microscope image of the HLC FPM, and the HLC Lyot stop.**

The Roman HLC coronagraph demonstrated $1.6 \times 10^{-9}$ contrast averaged from 3 to 9 λ/D over a 360 degree dark hole, using unpolarized light over a 10% bandwidth centered at 0.55 μm[56] (Fig. 12).

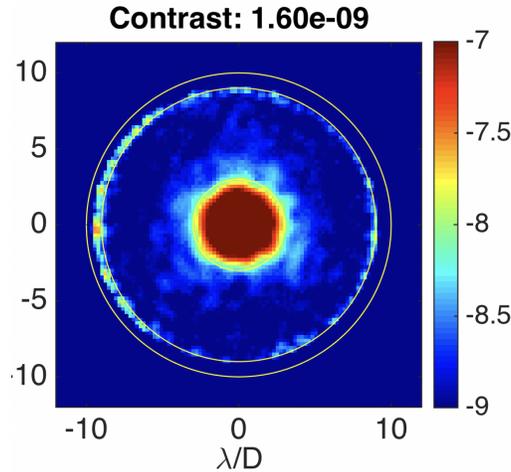

**Fig. 12. Best contrast measurements obtained in the lab with the Roman HLC coronagraph mask over 10% bandwidth (2019 technology milestone results[56])**

### 2.4.2 Roman SPC coronagraph

The shaped pupil coronagraph (SPC) principle[60,61] is to use optimum binary amplitude apodization of the entrance pupil to minimize diffraction effects in the subsequent focal plane. In the Roman case, the binary shaped pupil spatial distribution, focal plane mask (including a field stop) and Lyot stop designs were jointly optimized numerically[55]. As in the HLC case above, many performance metrics were considered, including spectral bandwidth, field of view, contrast, core throughput, encircled energy, DM surface height, and low-order aberration sensitivity. The layout of the "SPC-spec" coronagraph used for slit-spectroscopy of point sources with Roman is shown in Fig. 13. There are three such combinations of pupil apodizer, focal plane mask and Lyot stop, rotated by 60 degrees from each other to cover a full 360 dark degree dark hole.

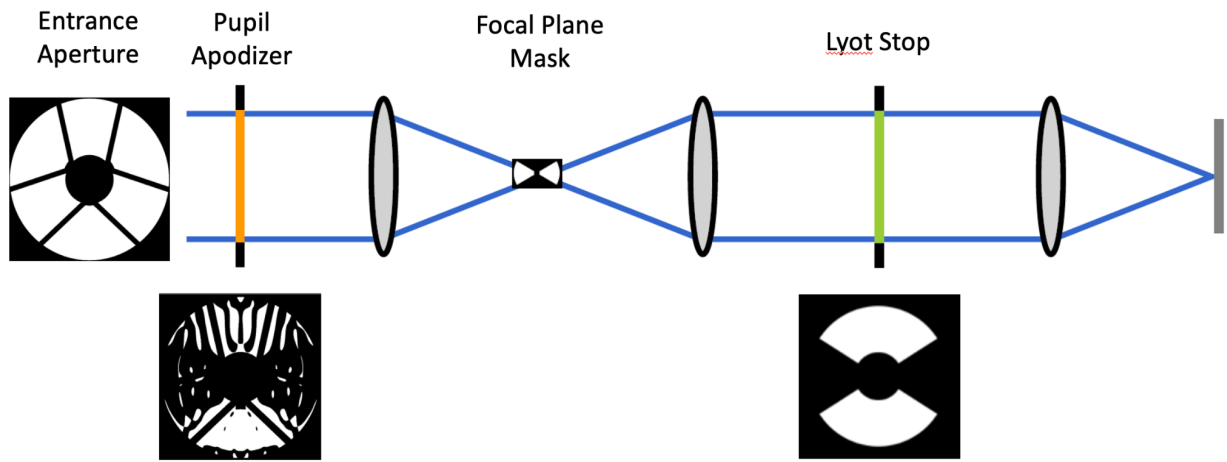

**Fig. 13. Optical layout of the Roman shaped pupil coronagraph dedicated to spectroscopic observations ("SPC-spec"), showing the Roman telescope entrance pupil, the binary shaped pupil used to apodize the pupil, the opaque (amplitude only) bow-tie shaped FPM, and the HLC Lyot stop.**

The Roman SPC coronagraph demonstrated 4.1 x 10⁻⁹ contrast averaged from 3 to 9 λ/D over a dark hole covering 2 x 65 degrees in azimuth, using unpolarized light over a 10% bandwidth centered at 0.55 μm[57] (Fig. 14). The same set-up reached 1.1 x 10⁻⁸ contrast over 18% bandwidth. The final filter bandwidth for spectroscopy with the flight coronagraph instrument was set at 15%. Because the SPC relies on amplitude apodization on the pupil, it is intrinsically less transmissive than coronagraphs using phase-only amplitude apodization or focal plane masks. Also, additional padding of the SPC mask along the spiders was necessary due to uncertainties in pupil alignment, distortion, and magnification. Both effects contributed to a low core throughput for the Roman SPC coronagraphs, similar to what was observed for the Roman HLC.

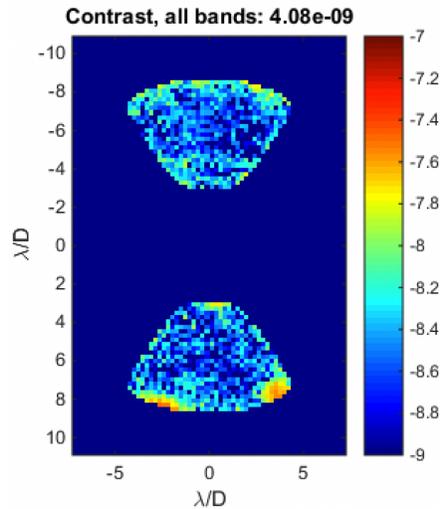

**Fig. 14: Best contrast measurements obtained in the lab with the Roman SPC-spec coronagraph over 10% bandwidth (2019 technology milestone results[57]).**

*2.5 Coronagraph Demonstrations for On-axis Segmented Telescopes*

*2.5.1 Phase-induced amplitude apodization complex mask coronagraph (PIAACMC)*

Although the ultimate theoretical performance of coronagraphs is not strongly dependent on aperture type[36], for on axis-telescopes, the secondary obscuration degrades the observed contrast performance of current coronagraphs. One way to mitigate this is with a variant of the classical PIAA coronagraph (Sec. 2.2.4) called the PIAA Complex-valued Mask Coronagraph (PIAACMC). In theory, a PIAACMC set-up can extend PIAA's intrinsic advantages in terms of inner working angle and throughput to any aperture shape[16,62] including on-axis segmented apertures. In addition to the two PIAA aspheric mirrors (PIAA optics), PIAACMC uses a reflective phase shifting focal plane mask, i.e., a special "complex valued mask" in place of the simple black spot occulter of the classical PIAA architecture. In practice, such an "ideal" mask can be implemented by a fully reflective holographic mask, which emulates the far-field response of the ideal CMC mask. In theory, the PIAACMC coronagraph with an ideal CMC mask and no manufacturing errors is capable of suppressing light from an on-axis point source for any arbitrary aperture in broadband light.

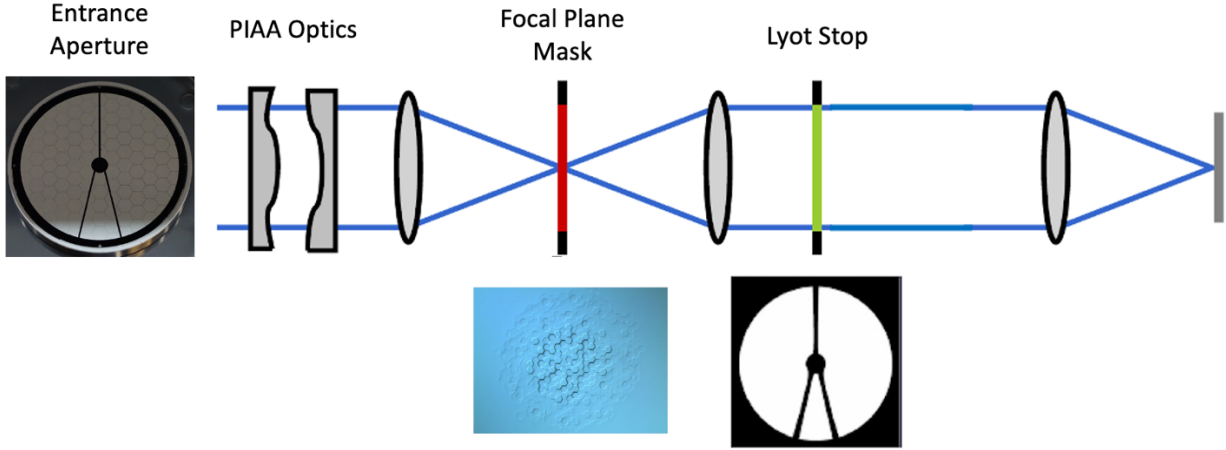

**Fig. 15. Layout of the PIAACMC design tested under vacuum in the HCIT. The entrance aperture, seen in reflection, emulates the on-axis segmented telescope proposed as part of the LUVOIR A mission study concept. A Boston Micromachines Corporation Kilo-DM (not shown) is conjugated to the second PIAA optic and performs wavefront control.**

The PIAACMC laboratory tests were conducted under vacuum, at the HCIT facility, producing the best contrast results so far for a *centrally obstructed segmented* aperture[63,64]. Key components of the layout are as follows (Fig. 15). The input light source is relayed to a pupil mask representing the LUVOIR A pupil (manufactured by Lambda Consulting using carbon nanotube technology) and then relayed to a PIAA system (by NuTek), consisting of a tube with two PIAA mirrors with a hole in the middle corresponding to the pupil obstruction. A Boston Micromachines Corporation Kilo-DM with 952 actuators is conjugated to the downstream PIAA mirror (PIAA2) and performs wavefront control. Finally, a CMC mask (by NASA JPL Micro Devices Lab), Lyot Stop (by Lambda Consulting), and field stop complete the coronagraphic setup. A polarizer (analyzer) was also placed in front of the camera to isolate polarization effects from other limiting factors. The testbed design was optimized for a one-sided (180º) dark zone using one DM. For sake of simplicity, and since the lab experiment aimed at demonstrating starlight suppression rather than off-axis point source detection, no "inverse PIAA optics" were used downstream of the Lyot stop to correct for the image distortion created off-axis by the PIAA optics. Such optics would need to be added for a PIAA based coronagraph instrument on HWO, adding a bit of complexity and extra losses.

The mean contrast performance obtained by the HCIT PIAACMC set-up between 3.5 and 8 λ/D in a 10% band was 1.9 x 10^{-8}, which is comparable (within roughly a factor of 2) to the monochromatic and 2% cases[63]. The 10% bandwidth case appeared to be limited by coherent chromatic effects.

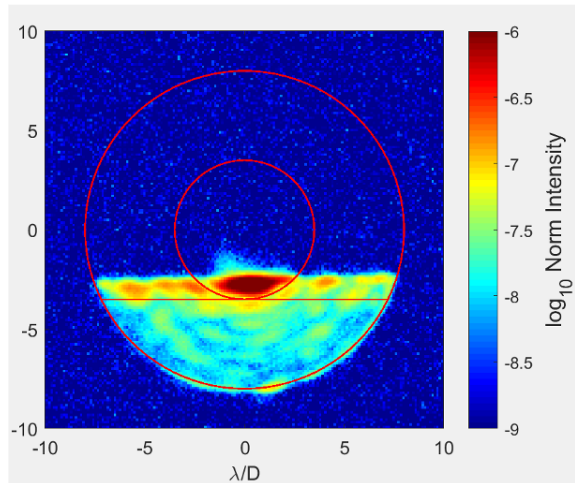

**Fig. 16. PIAACMC laboratory results: 10% bandwidth one-sided dark hole obtained around 0.65 μm. The mean contrast observed between 3.5 and 8 λ/D is 1.9 x 10⁻⁸. Adapted from[64].**

*2.6 Starshade Demonstrations*

In contrast to an (internal) coronagraph, where starlight suppression occurs by means of an optical system located after the telescope, a starshade, located tens of thousands of kilometers in front of the telescope along the line of sight to the target star, blocks starlight before it enters the telescope (Fig. 17). As a result, the contrast performance is independent of the type of telescope aperture used: on-axis or off-axis, monolithic or segmented.

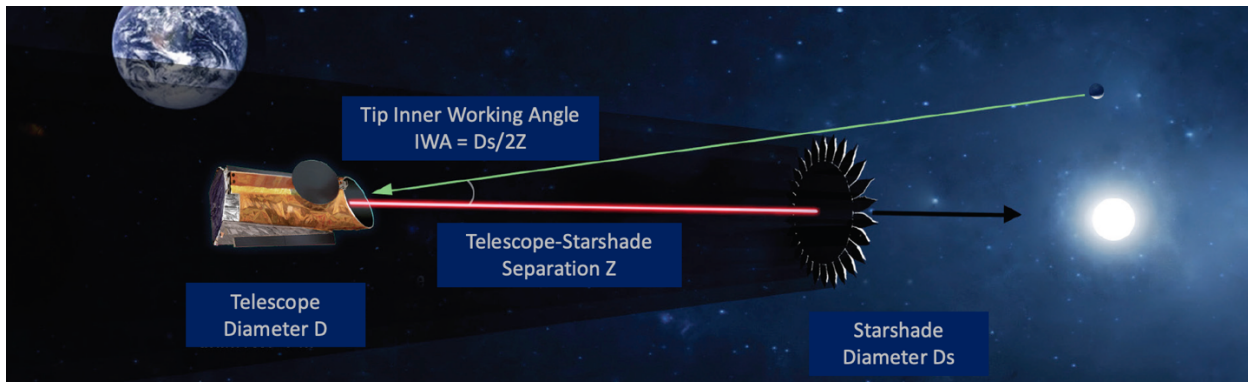

**Fig 17. Schematic exoplanet observations using an external starshade flying tens of thousands of kilometers in front of the telescope, along the line of sight to the target star (adapted from [2,8]).**

The starshade acts as a finite diffraction apodizer, and its shape is numerically optimized to minimize the amount of starlight diffracted across the telescope aperture over some user-specified wavelength range, potentially very broad, with designs theoretically providing better than $10^{-10}$ contrast over a bandwidth of 100% or more. A key starshade design parameter is its Fresnel number (F), defined at a given observing wavelength λ as $F = D_s^2/4\lambda z$, where $D_s$ is the starshade diameter and z its distance from the telescope. Sources located at angular separations greater than $D_s/2z$ are unaffected by the starshade, meaning that the starshade inner working angle (IWA) is of order $D_s/2z$ (although sources inside that angle are still partially transmitted). Adopting this definition of the starshade IWA, often referred as the "tip IWA", the starshade diameter required to reach a fixed IWA at a given wavelength is then $D_s = 2\lambda.F/IWA$. This means that the starshade diameter providing access to sources at a fixed *physical* angular separation is, to first order, independent of telescope diameter (D). Conversely, if the IWA is constrained to scale as $1/D$, e.g., if the desired IWA is $2\lambda/D$, the starshade diameter is simply F*D. For the flight applications considered here, Fresnel numbers of order 10 are considered a reasonable compromise between starshade manufacturability (e.g., the contrast sensitivity to shape errors which increases at small F numbers) and starshade size.

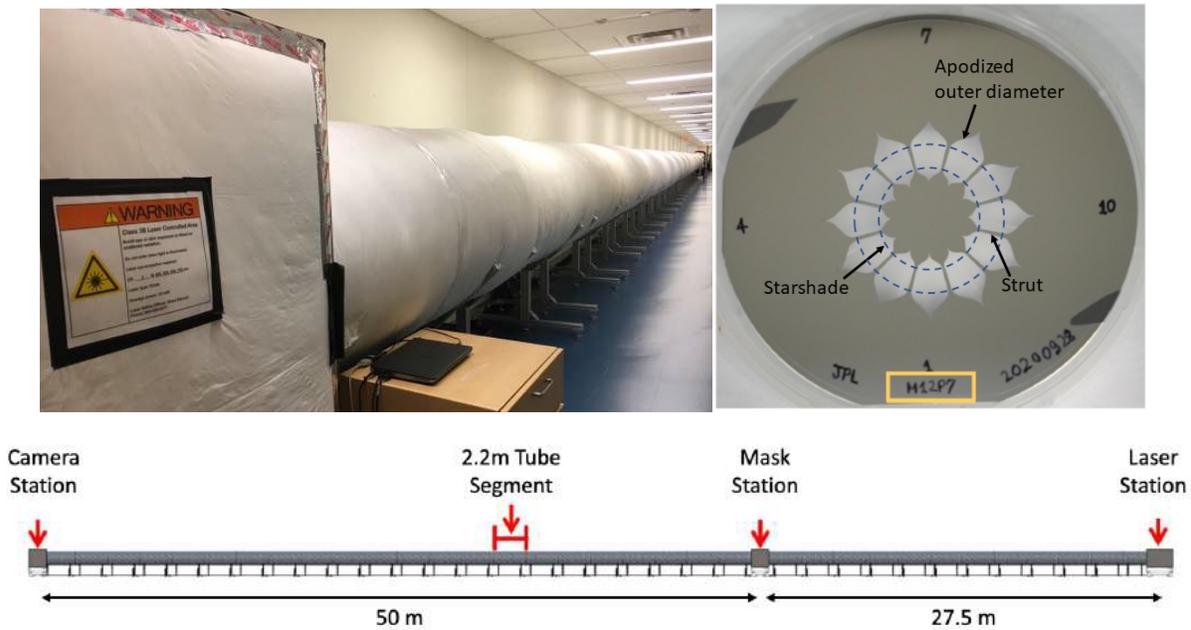

**Fig 18 (adapted from[65]). Top left: Standing at the camera station and looking down the length of the testbed toward the starshade mask and laser stations inside the Princeton Frick building. The beam propagates in air. The white insulation minimizes temperature variations and therefore turbulence in the tube. Top right: Starshade pattern etched into a silicon-on-insulator (SOI) wafer, manufactured at the Microdevices Lab at JPL. Interior to the inner blue circle is the inner starshade representing a free-floating occulter. The inner starshade is supported in the wafer by radial struts. The outer blue circle marks the start of the outer apodization function. Bottom: layout of testbed showing distances between laser, starshade, and camera.**

A miniature starshade mask (25.06 mm in diameter) was tested at the Princeton Starshade Testbed (S5 Milestone M1b[65]), which is 80 m long and capable of testing 1/1000th scale starshades at a flight-like Fresnel number. As shown in the schematic of Fig. 18, the beam from a spatially filtered laser propagates 27.5 m before it is diffracted by the starshade mask, after which it propagates another 50 m to a camera with an aperture diameter of 5 mm, sitting in the starshade's shadow. The beam line is enclosed in a 1 m diameter air-filled tube to seal the testbed from stray light and dust and to help stabilize the air inside. To emulate a broadband source, contrast measurements were obtained with a strongly (96%) linearly polarized laser source successively operating at four different wavelengths ranging from 0.641 to 0.725 µm, spanning an overall spectral bandwidth of 12%. The experiment Fresnel number[v] ranged from ~12.2 at 0.725 µm to ~13.8 at 0.641 µm, commensurate with flight values. As illustrated in Fig. 19 (right), the experiment demonstrated a broadband azimuthal mean contrast of ~ 2x $10^{-10}$ at a tip IWA of 1.7 λ/D at 0.725 µm (1.9 λ/D at 0.641 µm) and reached $10^{-10}$ contrast at ~ 2.1 λ/D for all wavelengths. The contrast performance further improved at larger angles and eventually was limited by Rayleigh scattering by air molecules[65,66]. Detailed modeling of non-scalar diffraction effects[66] indicates that the contrast performance and brighter lobes observed close to the IWA were limited by non-scalar diffraction (thick screen) effects, where the polarized light interacts with the edges of the mask. Non-scalar diffraction predicts that such effects will be completely negligible (> 1000 lower) on a much larger (> 10m) flight-size starshade. In the case of a much larger HWO-compatible starshade, which would have to be tens of meters in diameter, the main contrast limitations are instead expected to come from formation flying, deployment, petal shape and positioning accuracy, as well as solar glint.

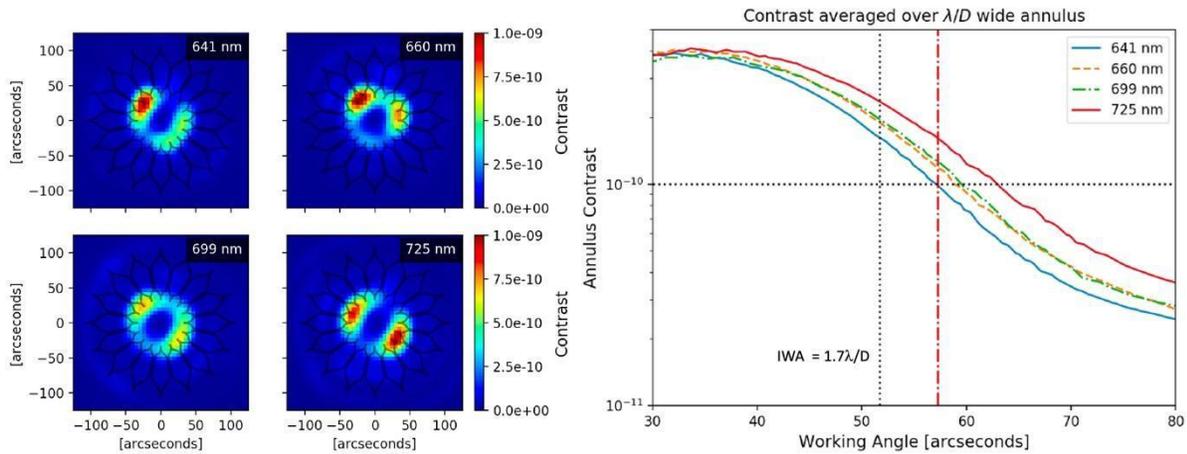

**Fig. 19 (adapted from[65]).** Left: 2D contrast (normalized intensity) curves measured on the Princeton testbed with the same starshade mask at 4 different wavelengths spanning 12% in bandwidth. The localized bright ~$10^{-9}$ lobes observed close to the starshade tip IWA are thick screen effects caused by interactions between the polarized source and the edges of the small mask. Right: azimuthal mean contrast vs. separation. The vertical black dotted line indicates the starshade mask IWA of 51.7", corresponding to 1.7 λ/D at 0.725 µm for the 5mm diameter aperture used. The separation where $10^{-10}$ contrast is reached corresponds to ~ 2.1 λ/D for all wavelengths considered.

---

[v] Given that the laser source is not at infinity, the effective starshade-telescope separation to be used for Fresnel number calculations[65] is 17.72m.

Figure 20 summarizes the contrast performance demonstrated in the lab as a function of angular separation for all starlight suppression systems considered above.

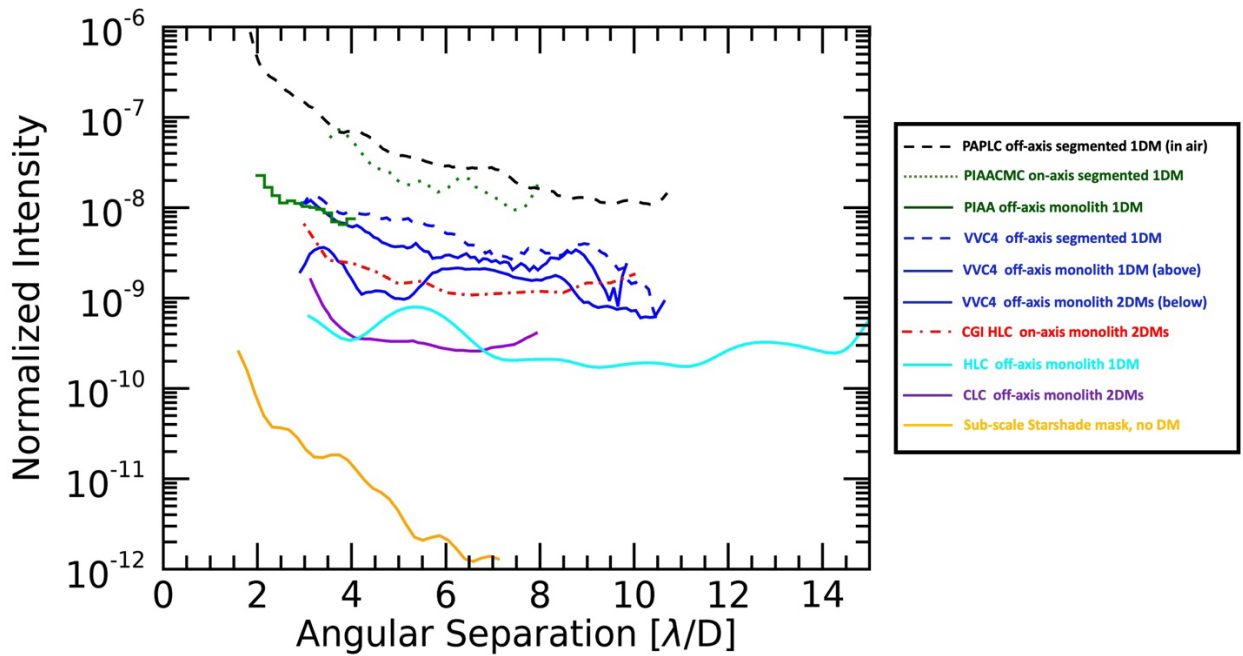

**Fig. 20. Best azimuthal mean contrast (normalized intensity) demonstrated to date by different starlight suppression approaches and laboratory experiments over a ~10% spectral bandwidth. The x-axis shows the angular separation in units of λ/D, where D is the entrance pupil *inscribed* diameter, λ is the central wavelength of the bandpass. Coronagraphic results were obtained with either one or two DMs and for different aperture types: off-axis monolith (plain curves); off-axis segmented (dashed curves); on-axis monolith (dashed dotted curve); and on-axis segmented (dotted curve). The sub-scale starshade results (orange curve) are independent of the aperture type considered. A summary of the experimental conditions used by the main starlight suppression demonstrations considered is given in Table 1.**

Before attempting to make any comparison, it is important to acknowledge that the lab experiments used different set-ups, as shown in Table 1 which summarizes the performance reached by each system and its main operating conditions: aperture type; DM type and format size; range of angular separations and azimuths over which the mean contrast is computed; central wavelength and bandwidth; polarization state; facility used for the demonstration; and vacuum vs. in-air operation. For the off-axis monolith case, Table 1 includes the CLC, HLC, and VVC4 results, but not the PIAA ones, which were significantly worse and obtained a small separation range. Also, the on-axis monolithic HLC and SPC Roman coronagraph results were presented above for completeness, but omitted in the summary table because the heavily obscured Roman entrance pupil does not represent a viable telescope aperture for HWO. Indeed, the HLC and SPC results with the Roman aperture have off-axis (planet) throughput values below 5% (Fig. 21), too low to spectrally characterize Earth-like exoplanets within reasonable exposure times (Sec. 4.4).

**Table 1. Summary of current broadband starlight suppression lab results and operating conditions. Results are expressed in units of normalized intensity (NI). CLC: Classical Lyot Coronagraph. HLC: Hybrid Lyot Coronagraph. VVC4: Vector Vortex Charge 4 Coronagraph. PAPLC: Phase-Apodized Pupil Lyot Coronagraph. PIAACMC: Phase-Induced Amplitude Apodization Complex Mask Coronagraph. A 64 x 64 DM was used for the HLC test ⁽*⁾, but the aperture diameter only extended over 48 actuators. In the CLC, VVC4 monolith and PAPLC cases, results are shown over ~10% and ~20-25% spectral bandwidths. The latter results are indicated *in italic*. All results were obtained in vacuum, except for the PAPLC and starshade mask. In the starshade case, starlight is canceled before the collecting aperture: results are independent of telescope aperture type and no active wavefront control is used (no DMs).**

| Lab Set-up | CLC | HLC | VVC4 | VVC4 | PAPLC | PIAA CMC | Starshade Mask |
|---|---|---|---|---|---|---|---|
| Aperture Type | **Off-axis monolith** | | | **Off-axis segmented** | | **On-axis segmented** | **Any** |
| Deformable Mirrors | 2 AOX 48x48 | 1 AOX 64x64* | 2 AOX 48x48 | 1 BMC 2k | 2 BMC 1k | 1 BMC 1k | None |
| Central wavelength (µm) | 0.550 | **0.800** | 0.635 | 0.635 | 0.660 | 0.650 | 0.680 |
| Spectral bandwidth | 10% *20%* | 10% *20%* | 10% *20%* | 10% | 9% *25%* | 10% | 12% |
| Number of polarizations | 1 | **2** | 1 | 1 | **2** | 1 | 1 |
| Dark Hole Separation Range | 3–8 λ/D *5–13 λ/D* | 3–15 λ/D | 3–10 λ/D | 3–10 λ/D | 2–13 λ/D *2–12 λ/D* | 3.5–8 λ/D | 1.7–7 λ/D |
| Dark Hole Extent | **Full** *One-side* | One-side | One-side | One-side | One-side | One-side | **Full** |
| Mean NI over Dark Hole | 4 x 10⁻¹⁰ *4 x 10⁻¹⁰* | 5.2 x 10⁻¹⁰ *1.8 x 10⁻⁹* | 1.6 x 10⁻⁹ *5.9 x 10⁻⁹* | 4.7 x 10⁻⁹ | 4.2 x 10⁻⁸ 9.5 x 10⁻⁸ | 1.8 x 10⁻⁸ | 2 x 10⁻¹¹ |
| NI at 3λ/D | 1.6 x 10⁻⁹ | 6.0 x 10⁻¹⁰ *2.3 x 10⁻⁹* | 2.4 x 10⁻⁹ *8.9 x 10⁻⁹* | 1.1 x 10⁻⁸ | 2.4 x 10⁻⁷ | ~7 x 10⁻⁸ | 2.1 x 10⁻¹¹ |
| Testbed Core Throughput at 3λ/D & 4.4λ/D | 0.08/0.13 | 0.10/0.19 | 0.38/0.46 | 0.38/0.46 | 0.51/0.53 | 0.60/0.61 | 0.68/0.68 |
| Facility and Testbed | HCIT-2 DST | HCIT | HCIT-2 DST | HCIT-2 DST | HiCAT | HCIT-2 | Princeton Frick |
| Vacuum Operation | Y | Y | Y | Y | **N** | Y | **N** |
| Main Reference | Seo et al. 2019[37] | Trauger et al. 2012[40] | Ruane et al.2022[49] | Riggs et al. 2022[53] | Por et al. 2020[54] | Marx et al. 2021[64] | Harness et al. 2019[65] |

For coronagraphs, the current best 10% bandwidth performance has been achieved on monolithic off-axis ("clear") apertures. Thus far, the only coronagraph architecture to have achieved better than $10^{-9}$ mean contrast at a separation smaller than $4\lambda/D$ over a full 360 deg dark hole is the CLC coronagraph operating behind a clear circular aperture. However, the CLC laboratory set-up used for that demonstration relied on a poorly-transmissive Lyot stop and had low (~8–13%) off-axis core throughput in the central 3 to 4.5 $\lambda/D$ region, likely too low for getting visible spectra of exoearths with a 6m HWO mission (Sec. 4). Next in terms of contrast performance is the linear HLC, which also provided better than $10^{-9}$ mean contrast over ~10% bandwidth but only over a one-sided dark hole. The linear HLC lab set-up offers slightly better core throughput (~10-20%) than the CLC coronagraph in the central 3 to 4.5 $\lambda/D$ region. The next best contrast performance, also achieved over a clear aperture and a one-sided dark hole, was obtained with the VVC4 coronagraph, which reached a contrast of a few $10^{-9}$ together with a significantly higher core throughput (~38%) at $3\lambda/D$.

In terms of segmented off-axis results, the only direct comparison comes from the broadband contrast performance obtained by the VVC4 on a segmented aperture with a single DM (blue dashed curve of Fig. 20). It is on average 1.5 x $10^{-9}$ higher than the monolithic 1DM case (upper solid blue curve).

This provides a first order of magnitude of the segmentation effect, but only at contrast levels of order a few $10^{-9}$, in a regime where the ultimate VVC4 broadband performance is still limited by residual spatial defects in the vortex mask rather than by the aperture type. To better assess the impact of segmentation itself, a more systematic comparison of vacuum results must be conducted with the same coronagraph set-ups, changing only the aperture from monolithic to segmented. This should include comparative tests of the CLC and HLC coronagraphs (only used so far on monolithic apertures), the PAPLC (only tested in air), as well as improved VVC masks recently manufactured.

Finally, the *on-axis* segmented aperture results (dashed green curve obtained with a PIAACMC) are currently significantly worse than the results obtained with the VVC4 on an *off-axis* segmented aperture and a single DM as well (blue dashed curve). However, it must be recognized that with only one vacuum coronagraphic experiment conducted so far, the testbed time devoted to on-axis segmented apertures demonstrations has also been significantly less than for off-axis apertures.

In terms of laboratory broadband starlight suppression results, the best contrast performance to date has been achieved with the starshade Princeton tested. The mean contrast demonstrated by the starshade mask at a flight-like Fresnel number was 2 x $10^{-11}$ over a 360 deg dark hole extending from 1.7 to 7 $\lambda/D$, with core throughput of 68%. At a separation of $2\lambda/D$ the contrast was 2 x $10^{-10}$, again with core throughput of 68%.

# 3 Beyond Raw Contrast

Given the extreme level of starlight suppression required for direct exoearth observations, laboratory experiments have mainly concentrated so far on demonstrating deep broadband (>~10%) raw contrast at small angular separations, as reported in the previous section. However, once the contrast performance is good enough that residual starlight levels fall significantly below the signal from irreducible astrophysical background sources (zodi and exozodi dust), the exposure time required to characterize exoearth planets becomes independent of raw contrast. As a result, the predicted yield increases marginally with further raw contrast improvements[32]. Depending on stellar distance and wavelength, that transition occurs at a threshold raw contrast level that ranges from ~$10^{-10}$ to ~$10^{-8}$ (Sec. 4.1).

Once that raw contrast threshold performance is reached, the number of exoearths characterized depends primarily on two starlight suppression KPPs: the overall optical system off-axis core throughput, especially its value at small angular separations, and the post-calibrated contrast uncertainty achieved after estimating and subtracting residual starlight speckles in raw science images. These KPPs are discussed in the following sections.

### 3.1 Core throughput

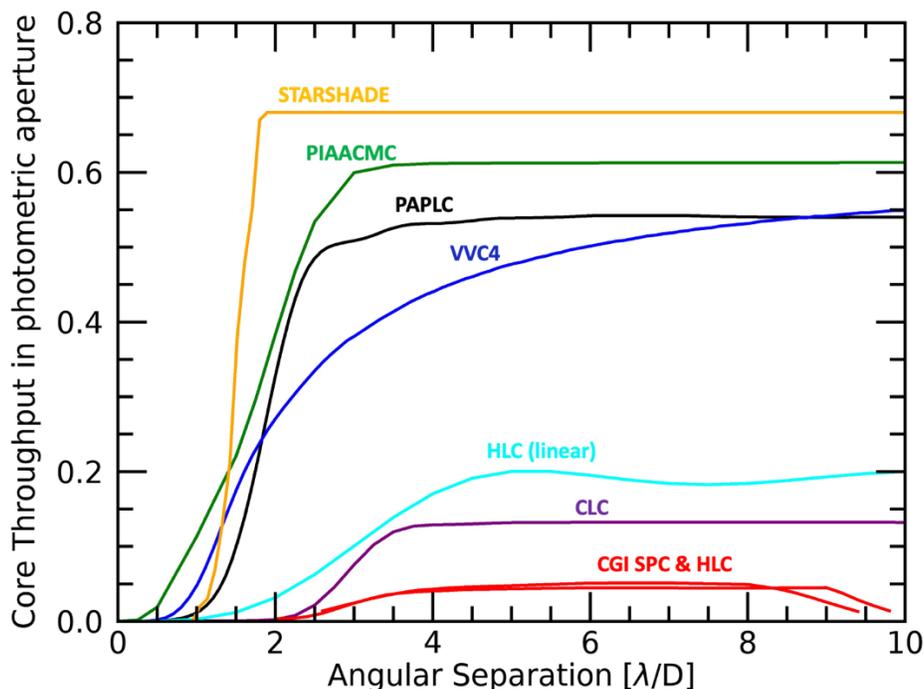

**Fig. 21: Core throughput of the laboratory starlight-suppression setups presented in Sec.2. Core throughputs are given as a function of angular separation in units of λ/D, where λ is the central wavelength and D is the inscribed diameter of the entrance aperture. At a given separation, the core throughput is computed within a circular aperture of radius 0.7λ/D for all systems, except for the Roman HLC and SPC-spec coronagraphs, which have highly-spatially-extended PSFs and for which the PSF FWHM region is used instead. In the starshade case, the angular separation is computed at 0.683 μm, the central wavelength of the Princeton 12% bandwidth lab set-up, and the core throughput is nearly achromatic vs. physical angular separation in arcsec. The PIAACMC curve assumes that inverse PIAA optics are used to correct for off-axis PSF distortion (although they were not part of the original lab set-up).**

The core throughput is the fraction of light from a point source at a given off-axis location that passes through the coronagraph or starshade masks and falls within a specified photometric aperture centered at that location. The photometric aperture can be chosen to optimize the detection of off-axis point sources, e.g., using a matched filter. The matched filter will depend on the telescope aperture type, the field-dependent PSF shape at the planet location and the background structure. It may provide significant sensitivity (SNR) gains over simple photometric apertures commonly used such as a circular aperture with a radius equal to the off-axis PSF FWHM or to a fixed value that maximizes the signal-to-noise ratio of an Airy pattern against a spatially uniform background[vi]. With two exceptions, all throughput values reported in this section use the *latter* approach and a constant circular photometric aperture of radius 0.7 $\lambda$/D, where D is the inscribed circular diameter of the telescope. The exceptions are the values for the CGI HLC and SPC-spec coronagraphs, which have bright PSF wings. In these two cases, the FWHM definition is used instead. The resulting photometric aperture radii are 0.57 $\lambda$/D for HLC and ~0.88 $\lambda$/D (averaged between the X and Y axes) for SPC-spec[58].

In the coronagraph case, the core throughput value captures transmission losses and off-axis PSF broadening due to the focal plane mask, Lyot stop, pupil amplitude apodization (if any), as well as the DM settings used to generate the dark hole. In the starshade case, the core throughput is approximately given by the product of the off-axis transmission of the laboratory mask by the fraction of the telescope PSF falling within the 0.7 $\lambda$/D radius photometric aperture (0.68 for a perfect Airy pattern).

Figure 21 shows the theoretical core throughput vs. angular separation of the tested starlight suppression systems presented in Section 2, under the experimental conditions and aperture types used in the laboratory. The Roman HLC and SPC coronagraphs exhibit significantly lower core throughputs than the other coronagraphs, not because these coronagraph types have intrinsically low core throughput, but rather because of the heavily obscured Roman telescope entrance pupil[58]. Their core throughput would be significantly higher on the less obscured aperture expected for HWO. The classical Lyot coronagraph, with full 360-degree dark hole and 10% bandwidth, used a Lyot stop blocking a substantial fraction of the aperture and is lowest in terms of core throughput performance for the HWO aperture. Depending on separation, it is 3 to 4 times lower than for the other coronagraph experiments considered, which are aimed at preserving high throughput at small separations. The PAPLC, PIAACMC (+ inverse PIAA inserted before science focal plane) and VVC4 all have > 20% off-axis core throughput values down to 2$\lambda$/D or less. The starshade core throughput curve is quite steep, and reaches its maximum value when the off-axis point source separation exceeds the angle sustained by the starshade when seen from the telescope, known as the "tip inner working angle (IWA)", which corresponds to ~1.8$\lambda$/D at the mean operating wavelength of 0.683 μm.

Comparing Fig. 20 and 21, the coronagraphs with the best contrast performance currently demonstrated in the lab (e.g., CLC and linear HLC) tend to have a lower off-axis throughput, while those with the best sensitivity to sources located at small separations (e.g., PIAACMC) tend to

---

[vi] In the case of a perfect circular beam of diameter d, the photometric aperture radius value that maximizes the signal-to-noise of an Airy pattern against a spatially uniform background is 0.665 $\lambda$/d. Assuming a Lyot stop whose diameter is 95% that of the telescope pupil of diameter D, this corresponds to 0.7$\lambda$/D.

have worse demonstrated contrast. Given that the HCIT essentially provides a static wavefront, this observed trend does not reflect the higher sensitivity to aberration drifts expected for coronagraphs with small inner working angle, which should only show up in dynamic testing. Yet, this small statistics trend is still observed.

At all separations larger than about 1.5λ/D, the theoretical core throughput is the highest in the starshade case, as there are neither occulter nor Lyot-plane losses inside the telescope. It reaches its maximum value of 68% (a perfect airy pattern) at the tip IWA.

## 3.2 Residual contrast after wavefront control and image processing

Dark hole contrasts will degrade in the presence of drifts. Active wavefront control for raw contrast stabilization and image calibration techniques have received more attention in recent years. We highlight hereafter some of these developments. As detailed in Sec. 4, the detection and spectral characterization of faint exoplanets require a combination of optical starlight suppression and image calibration (called "post-processing" in the coronagraph community) techniques. Depending on stellar distance, observing wavelength and raw contrast performance, planets of interest might be significantly dimmer than the residual starlight speckles and other background sources, such as the solar zodiacal light signal (spatially uniform over the dark hole) or the spatially varying exozodiacal signal in the target planetary system. Of specific interest is the accurate estimation and calibration of starlight residuals, which have the added complexity of varying over short timescales. As shown in[35], the relevant quantity is the spatial standard deviation of the starlight speckles *after* image calibration. Any residual error in the estimation of the speckles field in the science images will result in an irreducible systematic noise floor and a finite planet-to-star flux ratio detection limit, whatever the exposure time (Sec. 4.2). This error, and the resulting post-calibration contrast map residuals can be minimized in two complementary ways: (1) raw contrast stabilization using wavefront sensing and control (WFS&C) during the observations; and (2) accurate knowledge and subtraction of uncorrected residual speckles present in science images. The first method corrects for wavefront drifts in real time. It aims both at improving raw contrast over the science exposure and reducing the post-calibrated starlight residuals induced by raw contrast fluctuations. The second method is "after the fact", meaning that even if some speckle fluctuations cannot be perfectly corrected in real time (e.g. due to time lag or imperfect DM calibration), knowing what they were may further improve post-calibration residuals and get closer to the photon noise limit. We give hereafter some examples of speckle stabilization and speckle estimation techniques. We also acknowledge that a given WFS&C approach may both stabilize contrast and provide estimates of any uncorrected contrast fluctuations.

### 3.2.1 Estimating and correcting the electric field

Wavefront sensing and control is commonly used for ground-based observations to reduce the effect of turbulence and for laboratory deep starlight suppression demonstrations using coronagraphs. As very-high-contrast coronagraphic observations with HWO are concerned, an important goal is to limit non-common path errors between the wavefront sensor and the science beam path. This can be accomplished via direct measurements in the science focal plane or by picking up light reflected by the focal plane mask. In order to reach deep contrast, both low-order and high-order WFS&C are required, likely over different timescales. There have been numerous high-contrast WFS&C experiments worldwide, and we only highlight a few here to illustrate some approaches relevant to HWO, starting with low-order WFS&C.

One notable laboratory experiment is the active sensing and control of low-order aberrations ($Z_2$-$Z_{11}$: tip-tilt, defocus, astigmatism, coma, trefoil and spherical) at flight like photon flux demonstrated in vacuum for the Roman coronagraph[67]. In this case, the low-order WFS&C systems uses a low-order Zernike wavefront sensor (ZWFS), which picks up starlight reflecting off the metallic occulting spot of the focal plane mask, which is ~$6\lambda/D$ in diameter. The system was able to maintain coronagraph contrast below $10^{-8}$ with stability better than $10^{-9}$ over periods of up to an hour in the presence of purposely injected line of sight and low-order wavefront changes commensurate with the expected Roman space telescope on orbit jitter and thermal drift levels. Perturbations were injected at one DM, and then corrected at the other (located in a pupil plane conjugate) based on the ZWFS signal. This performance was demonstrated in both SPC and HLC modes, down to the photon flux expected from a V=2 star. At the lower flux expected from a V=5 star (closer to the magnitude of a typical HWO target), the low-order WFS&C system still suppressed the Roman-like line-of-sight disturbances ($Z_2$-$Z_3$) down to a post-correction jitter level of 0.35 mas rms[67]. Additional V=5 tests demonstrated the ability of the low-order WFS&C system to keep the 10-minute mean values of each low-order aberration ($Z_4$-$Z_{11}$) within 10 picometers over a period of 10 hours (B. Dube et al. in preparation). Wavefront stabilization over such timescales is in line with the Roman coronagraph operations concept, which baselines a series of 2-hr-long individual target star observations at alternating telescope rolls, together with higher-order touch-ups on a bright reference star every 10 hours or so. Low-order wavefront control of the first 11 Zernike modes using an FPM-filtered Zernike wavefront sensor was also demonstrated in air on the HiCAT testbed, using instead a classical Lyot coronagraph and a segmented aperture[68]. In that experiment, the in-air mean dark hole level was stabilized at a contrast level of 7 x$10^{-8}$ with a standard deviation of 7 x $10^{-9}$.

An inherent limitation of these experiments is that because the reflected starlight beam is spatially filtered by the small focal plane mask occulter, only low spatial frequencies (up to 3 or 4 cycles across the pupil) can be sensed by the ZWFS. Sensing residual phasing errors on a segmented telescope will require a WFS that is sensitive to higher spatial frequencies than that of the Roman coronagraph's WFS. HWO could have five or more mirror segments across the primary. Segment-to-segment errors in the form of rigid body motions will require at least a few wavefront samples per segmented mirror area, and at least two across to measure relative tilts. As a result, the focal plane mask reflective region needs to be more than ~20 $\lambda$/D in diameter to detect and correct such errors. If the WFS needs to measure DM instabilities at the actuator level, many more samples (potentially 48 to 128 actuators across) and a much larger reflecting region (respectively, >24 to 64 $\lambda$/D) will be required.

A first possible solution to sense these higher spatial frequencies at high sensitivity is to use out-of-band wavefront sensing, as is commonly done by ground-based adaptive optics systems. It was also proposed in the context of the LUVOIR study[69] where one of three parallel coronagraph channels, operating at different wavelengths, would be used for sensing the segment motions of the primary, while the other two are being used for science integrations. In that case, the ZWFS would be fed by a dichroic located upstream of the science coronagraphs, resulting in a fair amount of non-common path errors between the WFS and science channels. Other implementations have been suggested, including a dual purpose Lyot coronagraph (DPLC) mask for simultaneous high-contrast imaging and high spatial resolution wavefront sensing[70]. In this proposed approach, a reflective FPM is used to feed light to the WFS and minimize non-common path errors. The design uses a tiered metallic focal plane occulter to suppress the starlight transmitted through the mask at the science wavelength, and a dichroic-coated substrate that reflects out of band light to a ZWFS.

A second approach to sense and correct for high order wavefront drifts in real time is the so-called "dark zone maintenance" (DZM) algorithm[71]. In that case, the wavefront drift is estimated via dark hole intensity measurement while dithering of the deformable mirror actuators (similar in some sense to lock-in amplification). In addition, a predictive scheme based on an extended Kalman filter is used for maintaining the dark hole. This DZM approach has been validated in air in the HiCAT laboratory[72] at low star-equivalent flux levels with monochromatic 0.638 μm light. Over an annular dark zone extending from 5.8 to 9.8 λ/D, the closed-loop contrast was maintained at 5.3 x 10⁻⁸ with a standard deviation of 6.4 x 10⁻⁹, in the presence of a DM random walk drift of 20 pm rms per iteration (Fig. 22).

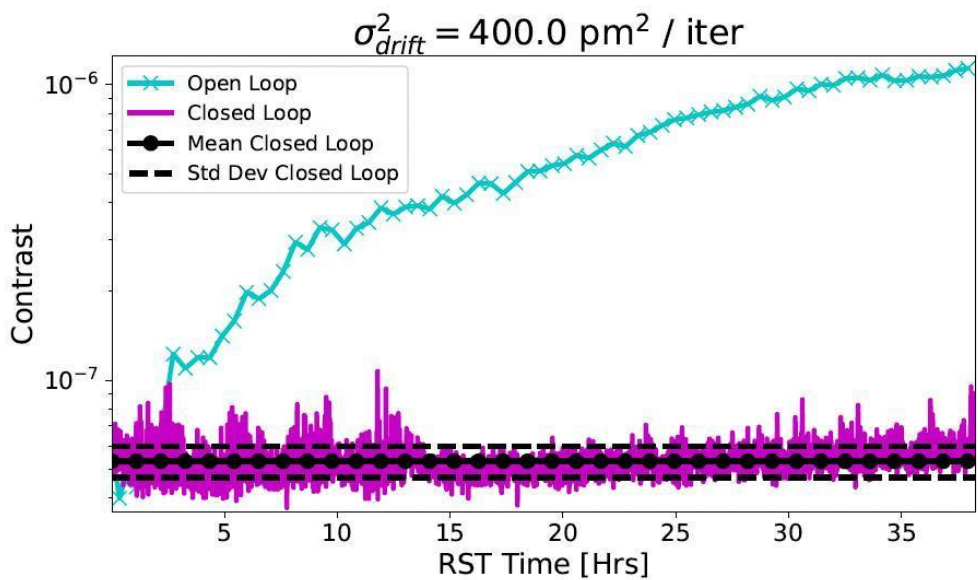

Fig 22 (from[72]). Mean dark-zone contrast measured vs time for the HiCAT low SNR dark zone maintenance experiment. In the presence of a DM random walk drift of 20 pm rms per iteration, the open-loop contrast diverges to 1.1 x 10⁻⁶ after 35 hours. Conversely, the closed-loop contrast is maintained at 5.3 x 10⁻⁸ with a standard deviation of 6.4 x 10⁻⁹. The iteration time is 39s.

A third approach to enable fast monitoring of wavefront changes at all spatial frequencies is linear dark field control (LDFC)[73,74]. In that case, there is no DM probing or dithering: the contrast in the dark hole is locked by monitoring the temporal evolution of bright speckles "outside" the dark hole, either spatially or spectrally (as in the DPLC). Changes in these bright field (BF) regions are highly correlated to the same wavefront changes that spoil the deep halo suppression in the dark field (DF). Because the BF images are significantly brighter than the DF images, they can be acquired at higher cadence, and no starlight needs to be directed to the DF during science exposures. By calibrating or computing the linear changes in the BF against wavefront changes, a linear dark field control (LDFC) servo can in principle maintain high contrast in the DF during science exposures. An initial demonstration was conducted in air with the Ames coronagraphic experiment testbed[75,76] using spatial LDFC (i.e., bright uncorrected regions) to stabilize a 5x10$^{-7}$ dark hole extending from 1.5 to 5.2 $\lambda$/D in the presence of injected perturbations.

An interesting upcoming trade for WFS&C in the context of HWO's coronagraphs is then between the LDFC approach using bright light outside the dark hole, the out-of-band high-order DPLC-like ZWFS approach, and a more conventional (CGI-like) low-order in-band ZWFS working together with a (DZM-like) DM-dithering scheme with sensing in the science dark hole. Testing of all three approaches using a common testbed and coronagraph mask would be most informative.

### 3.2.2 Estimating and subtracting residual speckles

Whatever WFS scheme is used to minimize the instrumental raw contrast and its fluctuations, a residual starlight speckle field will remain at some level in the science images. For optimal exoplanet detection, the raw contrast should be deep enough that residual starlight is fainter than the solar zodi plus exozodi background. The speckle field must then be estimated and subtracted accurately enough to get back to that background shot noise limit. For a given science image, the speckle field may be estimated using a library of coronagraphic images obtained at different times on the same target (but at a different roll angle of the telescope) or on reference targets. This is the principle of, e.g., the successful LOCI[77] and KLIP[78] high contrast post-processing approaches commonly applied to reference (RDI) and angular differential imaging (ADI) of coronagraphic observations down to detection limits of $\sim 10^{-6}$, from the ground and with HST/JWST. However, these are "blind" correction methods, where the speckle field at the time of science observations is estimated through observations obtained at a different time, and possibly on a different star (RDI). Accordingly, such methods place strong requirements on wavefront and speckle temporal stability during telescope slews (RDI) or rolls (ADI).

A first alternative is to leverage auxiliary wavefront sensing data to estimate the starlight electric field at the time of the science observations[79]. In particular, if bright fields recorded at other wavelengths (e.g., DPLC WFS) or in different regions (LDFC WFS) of the science focal plane can indeed be used to reliably estimate the starlight residuals in the dark hole, then high-accuracy estimates of the speckle field can be made over much shorter timescales than when using the dark field science images themselves. If such speckle "self-calibration" estimates[80] can be made faster than the speckle field changes, wavefront fluctuations will be accurately known. Even if the DMs cannot perfectly correct for these measured fluctuations in real time due to finite spatio-temporal response or chromatic effects, such fluctuations may be corrected after the fact in science images, potentially resulting in a significant relaxation of wavefront stability requirements. In this case, raw contrast stability needs only to be maintained at a level below the (exo)-zodi background, and no longer at a fraction of the planetary signal. As long as this condition is met, this essentially becomes a wavefront knowledge problem. Importantly, this scheme requires that the relationship between the WFS bright field data and the dark field is constant over time, or at least that any time variable offset is slow enough to be measured. Machine learning algorithms trained to estimate the electric field in the dark field based on bright field measurements are being investigated. Post-processing schemes using WFS information directly gathered in the dark hole at the science wavelength would likely suffer less from that limitation and have been suggested as well, for instance using DM dither and reduced order estimation of the speckle electric field[81,82].

A second alternative is the use of medium to high resolution (R > 1,000) observations to distinguish residual starlight and (exo)-zodi background signals, essentially featureless or showing slow spectral variations, from planetary (or low-mass companion) signals that show distinctive molecular absorption features at specific wavelengths and exhibit line forests over the science bandpass. In this approach, major atmospheric species can be searched for one by one in the companion's atmosphere by fitting the spectra obtained with model templates of individual molecules (the "molecular mapping" technique). Initial demonstrations were obtained at spectral resolution high enough to leverage the planet differential radial velocity relative to the star, resulting in the direct detections of young extrasolar giant planets with integral field spectrographs at both VLT[83-84] and Keck[85,87], and providing direct measurements of planet molecular content, as well as temperature, surface gravity, radial and spin velocities. Whether this approach might be applicable to the significantly fainter ($10^{-10}$ vs $10^{-5}$) planets to be spectrally characterized by HWO remains to be investigated. Previous studies of such high dispersion coronagraphy in the case of the HabEx and LUVOIR designs[88,89] suggested that a spectral resolution of ~1000 might be optimum. However, this value strongly depends on wavelength, detector properties, and overall system throughput, and should be reassessed in the HWO context. We also note that this method recently demonstrated the detection of a ~$10^{-5}$ close-in brown dwarf companion with JWST using medium resolution (R~2700) NIRSpec IFU data[90] and is also being developed at Keck in conjunction with a coronagraph for the first time, using a single-mode fiber to feed a high-spectral-resolution (R~35,000) point source spectrograph[91,92].

Overall, the current developments in WFS&C, starlight speckle estimation and spectral disambiguation point to two major upcoming trades in the context of HWO coronagraph design and requirements. The first trade is between speckle stability and speckle knowledge, assessing how far stability requirements might be realistically relaxed. The second one is about spectral resolution and detector noise, assessing how an increase in spectral resolution (wrt the R~100 value commonly assumed) might be possible, and what benefits it might bring.

## 4 Illustrative Cases

In the previous sections, we presented the current state of the art in terms of raw contrast demonstrated in the lab (Sec. 2) and introduced the other two main characteristics of a starlight-suppression system: off-axis (core) throughput and post-calibration contrast (Sec. 3).

This conceptual framework can now be applied to start assessing how the currently demonstrated lab performance compares to the notional needs of HWO. HWO's specific exoplanet science figures of merit, characterization requirements, and candidate architectures are not defined yet, making complete design reference mission simulations premature. Instead, we concentrate here on a fiducial exoearth twin case and estimate the exposure time required to detect and spectrally characterize it for different combinations of starlight suppression system key performance KPPs. This approach using exposure time as a basic cost function already enables a rough exploration of the starlight suppression performance needed by HWO, and the identification of several important design trades.

## 4.1 Astrophysical Signals

In this section, we aim to compare the signals coming from the solar zodiacal light, the exozodiacal light, the planet light and residual starlight. Following Stark et al.[31], we adopt adopt a mean surface brightness of the solar zodi in the V band of 23 mag/arcsec$^2$, and a surface brightness of 22 mag/arcsec$^2$ for one zodi worth of exozodi dust located in the habitable zone of a Sunlike star. Indeed, a solar zodi cloud analog would appear brighter when observed from afar, as we would typically observe dust both above and below the mid-plane, and the $1/r^2$ illumination factor will always bias the background flux in any photometric aperture to higher values.

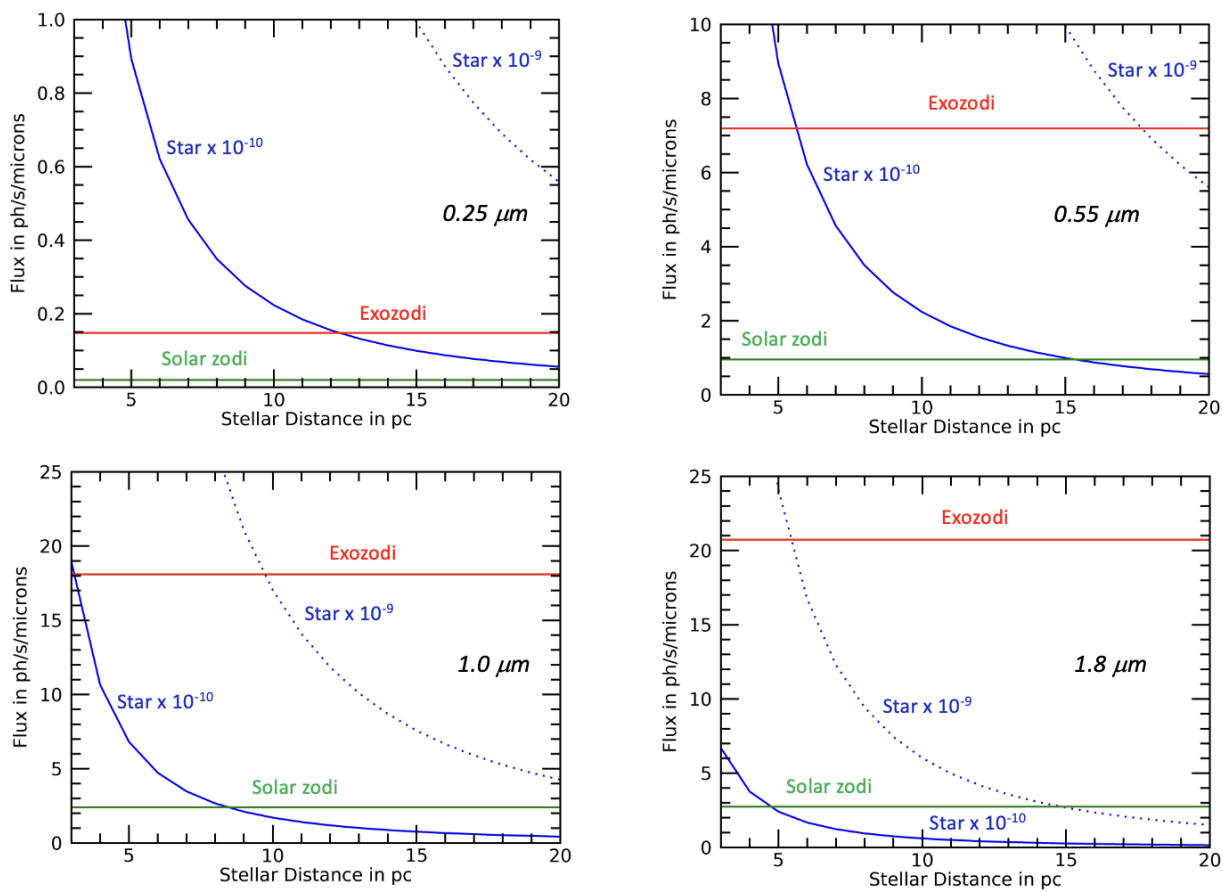

**Fig. 23.** Astrophysical fluxes detected at 1 AU from a Sun-like (G2V) star, in a photometric region of radius 0.7λ/D, as a function of stellar distance. The four panels correspond to wavelengths of 0.25, 0.55, 1.0, and 1.8 μm (note the different y-axis ranges). All cases assume a D=6m telescope with a starlight suppression system with unit end-to-end transmission (including detector) efficiency, and an off-axis PSF described by a perfect Airy pattern (i.e., with a core throughput of 0.68 within the photometric region). The starlight suppression raw contrast at 1AU is set to either $10^{-10}$ (plain blue curve) or $10^{-9}$ (dotted blue curve). The blue curves also correspond to the flux from an exoplanet with a planet-to-star flux ratio of $10^{-10}$ (an exoearth at quadrature) or $10^{-9}$. The exozodi level is set to 3 zodis in all cases.

Figure 23 shows the "incident" astrophysical fluxes detected at 1 AU from a Sun-like (G2V) star, in a photometric region of radius 0.7 λ/D, as a function of stellar distance, at wavelengths of 0.25, 0.55, 1.0, and 1.8 μm. All cases assume a 6m (inscribed) diameter telescope with a starlight suppression system with unit end-to-end transmission (including detector) efficiency, and an off-axis PSF described by a perfect Airy pattern (i.e., with a core throughput of 0.68 within the prescribed photometric region[vii]). The starlight-suppression raw contrast at 1AU is set to either $10^{-10}$ (plain blue curve) or $10^{-9}$ (dotted blue curve). Because of the way raw contrast is defined, the plain blue curve also corresponds to the flux from an exoplanet with a planet-to-star flux ratio of $10^{-10}$ (an exoearth at quadrature), and the dotted blue curve corresponds to a planet flux ratio of $10^{-9}$. The exozodi level is set to 3 zodis in all cases, the most likely median exozodi level derived from the LBTI HOSTS survey[93]. As a result, the assumed exozodi V-band surface brightness in the habitable zone is 20.8 mag/arcsec$^2$. Because the solar zodi and exozodi sources are spatially extended, their corresponding signal within the (0.7 λ/D) photometric aperture is constant vs stellar distance but increases with wavelength. Conversely, the residual starlight signal decreases with stellar distance, meaning that observations will become dominated by "background" (exo)-zodi signals for distant enough targets and/or deep enough starlight cancellation.

At V-band (around 0.55 μm, top-right plot), assuming an instrumental raw contrast level of $10^{-10}$, the (3 zodis) exozodi signal dominates over starlight residuals, or equivalently the signal from a $10^{-10}$ Earth-like planet, for all stars further away than 6 pc, i.e, most of the targets. This says that at V-band, unless most stars have significantly less than 3 zodis worth of dust, a raw contrast of $10^{-10}$ or better will guarantee that most observations are limited by the exozodi background rather than starlight residuals, at least if the latter are constant or perfectly calibrated. The irreducible solar system zodi signal only dominates for stars further than ~15 pc. On the other hand, a raw contrast of $10^{-9}$ at V-band results in starlight residuals being the dominant source of photon noise for all Sunlike stars within ~17 pc, likely most of the sample. However, if most stars have significantly more than 3 zodis worth of dust, a high exozodi scenario still compatible with the LBTI exozodi survey data which derived a 95% confidence upper limit of 27 zodis on the median exozodi level of Sunlike stars, exozodiacal light will still dominate. Depending on star distance, a raw contrast between $10^{-10}$ and $10^{-9}$ will hence be required to guarantee that background (exo-)zodi signals dominate over stellar short noise.

At shorter wavelengths, e.g., in the near UV around 0.25 μm (top-left plot), the solar and exozodi backgrounds are both reduced due to the smaller beam and photometric region size compared to V-band. A raw contrast of $10^{-10}$ or lower is required to remain exozodi-background-limited for near UV observations of stars closer than ~12 pc. The raw contrast requirements are hence most stringent at near UV wavelengths.

At longer wavelengths of 1μm (close to the strong 0.94 um water band) and 1.8 μm, we see the opposite effect. Most observations would still be limited by the solar and exozodi backgrounds, even for raw contrast as bad as $10^{-9}$ or even close to $10^{-8}$ (bottom plots of Fig. 23). The raw contrast requirements are in principle least stringent at the longer infrared wavelengths. However, we

---

[vii] If the off-axis PSF is more extended, the stellar and planetary fluxes per photometric aperture will decrease accordingly. Conversely, the solar and exozodi fluxes per photometric aperture will remain essentially constant.

caution that operating at a raw contrast of 10⁻⁹ or higher will make the calibration of residual speckles below the planet level more challenging, as it will require speckle calibration at even better relative precision to maintain the same noise floor. Exoearth observations at near infrared wavelengths would also put a premium on the precise estimation of the exozodi signal at the planet location. Indeed, at 1.8 μm 3 zodis worth of exozodi dust would create a signal ~50x brighter than an exoearth seen at quadrature around a star located at 12 pc.

It is worth highlighting that the level of exozodiacal light around sunlike stars has been best estimated in the mid-infrared, down to an uncertainty of ~50 zodis (1σ) per *individual* star, assuming a solar-like grain size and density profile. It is much less constrained at HWO's near-UV to near-IR wavelengths. As shown in Fig. 23, even at only 3x the solar system level, exozodi dust represents the dominant source of background shot noise for most exoearth observations at visible to near infrared wavelengths. At 10–20x the solar density level, exozodi dust clouds may also create confounding resonant structures[94]. The exozodi surface brightness - and its possible time variability[95,96] - has then a large impact on the observability of exoearths around *individual* targets. Precursor observations at or close to HWO's wavelengths would thus be very valuable and should be conducted. Among them, an exozodi survey conducted by the Roman coronagraph at visible wavelengths would be particularly impactful[97-99].

### 4.2 Signal-to-noise ratio (SNR) and time to SNR

For signal-to-noise calculations, we adopt the definitions and formalism described in previous work[35]. Over exposure time δt, for a given spectral bandpass, and assuming no systematic noise (photon noise only) the planet observation signal-to-noise ratio is given by:

$$SNR = \frac{r_{pl} \cdot \delta t}{\sqrt{r_n \cdot \delta t}} \quad , \quad (1)$$

where $r_{pl}$ is the planet photon rate detected in a circular photometric aperture centered around the planet location, and $r_n$ is the total photon rate detected from all astrophysical sources[viii] in the aperture: planet ($r_{pl}$), residual starlight (speckle rate $r_{sp}$), solar zodiacal light ($r_{sz}$) and exozodiacal light ($r_{xz}$). By definition:

$$r_n = r_{pl} + r_{sp} + r_{sz} + r_{xz} . \quad (2)$$

Following detailed calculations presented in[100], we have:

$$r_{pl} = \eta_p . \varepsilon . N_s , \quad (3)$$
$$\text{and } r_{sp} = \eta_p . RC . N_s , \quad (4)$$

where $\eta_p$ is the field-dependent point source core throughput computed at the planet location (x,y) over the specified photometric aperture, as defined in Sec. 3.1. Following[35], it is defined as

---

[viii] Strictly speaking, the planet photon noise contribution should only be included for a characterization (i.e., a flux measurement) SNR and not for a detection SNR.

the product of the overall occulter transmission $\eta_{occ}$ (x,y) (resulting from all coronagraphic masks or the external starshade) and the fraction of light in the planet PSF that ends up in the photometric aperture $\eta_{PSF}$ (x,y), so that

$$\eta_p(x,y) = \eta_{occ}(x,y).\eta_{PSF}(x,y) . \quad (5)$$

In the starshade case, $\eta_{occ}$ (x,y) is the off-axis transmission profile of the external occulter, reaching ~1 at the starshade petal tip. In the coronagraph case, the transmission of the focal plane mask $\eta_{FPM}(x,y)$ is multiplied by the constant transmission of the Lyot stop ($\eta_{LS}$) and of any pupil plane mask apodizer ($\eta_{PPM}$), so that

$$\eta_{occ}(x,y) = \eta_{PPM}.\eta_{LS}.\eta_{FPM}(x,y) . \quad (6)$$

For uniform extended sources that are spatially uniform over larger scales than the specified photometric aperture, such as zodiacal light and exozodi to a high degree, there are no PSF losses and only the $\eta_{occ}$ factor applies, so that the equivalent throughput factor accounting for (exo)-zodi transmission through the starlight suppression system is given to first order by

$$\eta_{sz}(x,y) = \eta_{xz}(x,y) = \eta_{occ}(x,y) = \eta_p(x,y)/\eta_{PSF}(x,y) , \quad (7)$$

where
- $\varepsilon$ is the mean astrophysical (intrinsic) planet-to-star flux ratio over the bandpass,
- RC is the starlight suppression system "raw contrast" at the planet location, defined in Sec. 2.1,
- $N_s$ is the total number of stellar photons that would be detected per second through the optical system over some bandpass $\Delta\lambda$ in the absence of any coronagraphic (pupil and/or focal plane) masks, Lyot stop or starshade mask, integrating starlight received over the whole focal plane. That is[100]:

$$N_s = \int_{\Delta\lambda} \phi_s(\lambda) A. q(\lambda).Tr(\lambda) . d\lambda , \quad (8)$$

where $\phi_s(\lambda)$ is the incoming stellar flux in photons per unit area per unit time per unit wavelength at the primary mirror, A is the collecting area of the telescope, q($\lambda$) is the detector finite quantum efficiency, and Tr($\lambda$) is the overall transmission and reflectivity of all *non-coronagraphic* optics in the system which affects the star, planet and (exo-)zodi light equally, including any spectral filters and polarizers.

As already mentioned, the planet signal will be accompanied by a background signal made up of three components: residual starlight, solar zodi and exozodi, some of which could be significantly brighter than the detected signal from an exoearth depending on stellar distance and observing wavelength (Sec. 4.1). Some form of differential imaging based on advanced post-processing techniques such as reference, angular, spectral, polarization, coherent differential imaging or wavefront sensor-based estimates will be required to properly estimate the background and subtract it. That background subtraction will generally be imperfect, meaning that the planet SNR will include a systematic post-calibration error term and can be rewritten as:

$$SNR = \frac{r_{pl} \cdot \delta t}{\sqrt{r_n \cdot \delta t + r_{\Delta I}^2 \cdot \delta t^2}} \qquad (9)$$

The systematic noise term growing as $r_{\Delta I} \cdot \delta t$ captures the finite ability to calibrate the total background signal at the planet location after using post-processing. Following[35] and extending their approach to all three background sources, $r_{\Delta I}$ can be thought of as the residual total background rate after post-processing and written as:

$$r_{\Delta I} = f_{\Delta I} \cdot r_{sp} + f_{\Delta Isz} \cdot r_{sz} + f_{\Delta Ixz} \cdot r_{xz} \qquad (10)$$

where $f_{\Delta I}$ (<1) is the effectiveness of the starlight differential imaging suppression. Similarly, the $f_{\Delta Isz}$ and $f_{\Delta Ixz}$ coefficients measure the ability to reduce the effective solar and exozodi flux after image calibration. The solar zodi signal is spatially uniform in the dark hole and constant over time to first order, making it relatively straightforward to calibrate precisely. Conversely, the exozodi signal depends on the target star itself. It will be estimated from the science images, through modeling of the exozodi brightness distribution and its intensity at the planet location. In the remainder of this document, we assume that the zodi and exozodi signals are perfectly calibrated, i.e., that $f_{\Delta Isz} = f_{\Delta Ixz} = 0$. This is clearly an optimistic assumption for exozodi, especially at longer wavelengths, where exozodi may get significantly brighter than planet light (Sec 4.1). Regardless, we assume hereafter that the only systematic noise term remaining comes from imperfect subtraction of starlight speckle residuals and their temporal fluctuations during science exposures, so that:

$$r_{\Delta I} = f_{\Delta I} \cdot r_{sp} \qquad (11)$$

In the case where the speckle background mean value can be perfectly removed, $f_{\Delta I} = 0$ and the observations are purely shot noise limited. As a best-case scenario, we further assume zero detector noise in all subsequent SNR calculations.

Based on Equations (1) through (11), and assuming that differential imaging doubles the shot noise term (e.g., as in the case of angular differential imaging), the planet detection SNR can be rewritten as:

$$SNR = \frac{\eta_p \cdot \varepsilon \cdot N_s \cdot \delta t}{\sqrt{2 \cdot (\eta_p \cdot \varepsilon \cdot N_s + \eta_p \cdot RC \cdot N_s + r_{sz} + r_{xz}) \cdot \delta t + (f_{\Delta I} \cdot \eta_p \cdot RC \cdot N_s \cdot \delta t)^2}} \quad . \quad (12)$$

If differential imaging is based on the observations of a brighter reference star or on simultaneous observations of the target star (Sec. 3.2), the shot noise penalty will be lower, so that Eq (12) describes the worst-case impact of post-processing on SNR and exposure time required.

Noting that $f_{\Delta l}$ corresponds physically to the ratio of the post-calibrated contrast spatial rms at the planet location, noted $\sigma_{\Delta C}$ hereafter, to the raw contrast RC at the planet location[ix], the SNR can be rewritten as:

$$SNR = \frac{\eta_p.\,\varepsilon.\,N_s.\,\delta t}{\sqrt{2.\,(\eta_p.\,\varepsilon.\,N_s + \eta_p.\,RC.\,N_s + r_{sz} + r_{xz}\,).\,\delta t + (\eta_p \sigma_{\Delta C}.\,N_s.\,\delta t)^2}} \quad . \quad (13)$$

For a given astrophysical scene and non-coronagraphic end-to-end optical transmission, the achievable SNR hence depends on 3 starlight suppression parameters computed in the photometric aperture centered at the planet location: the off-axis core throughput ($\eta_p$), the raw contrast (RC), and the post calibrated contrast $\sigma_{\Delta C}$. For exposure times long enough that systematic effects prevail over photon noise, the achievable SNR approaches an asymptotic maximum value given by:

$$SNR = \frac{\varepsilon}{f_{\Delta l}.\,RC} \; = \; \frac{\varepsilon}{\sigma_{\Delta C}} \qquad (14)$$

This means that the minimum planet-to-star flux ratio $\varepsilon$ detectable with an SNR of 1 after calibration of the raw images using differential imaging post-processing techniques is equal to $\sigma_{\Delta C}$ (Sec. 3.2). And whatever the exposure time, if the characterization SNR threshold is set to a value $SNR_0$, no planet can be successfully observed with a planet-to-star flux ratio lower than

$$\varepsilon_{min} = SNR_0.\,f_{\Delta l}.\,RC \; = \; SNR_0.\,\sigma_{\Delta C} \quad . \quad (15)$$

As an illustration, to spectrally characterize a planet with a flux ratio of $\varepsilon = 10^{-10}$ with an SNR of 10, the post-calibrated contrast rms $\sigma_{\Delta C}$ must be lower than $10^{-11}$. And for all planets brighter than $\varepsilon_{min}$, the exposure time required to reach the threshold SNR is:

$$t_{SNR_0} = \frac{2.\,SNR_0^2.\,(\eta_p.\,\varepsilon.\,N_s + \eta_p.\,RC.\,N_s + r_{sz} + r_{xz})}{(\eta_p.\,\varepsilon.\,N_s)^2 \; - SNR_0^2.\,(\eta_p.\,\sigma_{\Delta C}.\,N_s)^2} \qquad (16)$$

While the time to reach a given SNR always decreases with improved raw contrast, all other parameters being fixed, the minimum planet flux ratio accessible (Eq. 7) is set by the post-calibration contrast rms $\sigma_{\Delta C}$ defined as the product of raw contrast (RC) and speckle post-processing efficiency ($f_{\Delta l}$). This means that both terms must be considered and minimized through the instrument design and operations concept. It also provides the opportunity to trade one term for the other, at least within the range of raw contrast values that preserve a reasonably short exposure time (Sec. 4.3 & 4.4).

In summary, we made three strong assumptions in the above SNR calculations. Two are optimistic and one is conservative. On one hand, we assumed negligible detector noise, which may be overly

---

optimistic depending on the spectral resolution used, as well as "perfect" subtraction (down to the photon noise limit) of the (exo-)zodi background at the planet location. Both are best case scenarios. On the other hand, we assumed an ADI-like post-processing approach that doubles the background shot noise and observing time, which is a worst-case scenario.

In the case where there is no systematic stellar noise ($\sigma_{\Delta C} = 0$) and the exozodi background flux dominates over the residual stellar flux, we have:

$$t_{SNR_0} \propto \frac{SNR_0^2 \cdot r_{xz}}{\left(\eta_p \cdot \varepsilon \cdot N_s\right)^2} \quad . \quad (17)$$

Because the exozodi background is spatially extended, its detected flux does not depend on stellar distance or telescope diameter and scales as

$$r_{xz} \propto \frac{\eta_{occ} \cdot Tr}{R} \quad . \quad (18)$$

Using Eq (5) and (8), we get:

$$\eta_p \cdot N_s \propto \frac{\eta_{occ} \cdot \eta_{PSF} \cdot Tr \cdot D^2}{R \cdot Dist^2} \quad , \quad (19)$$

where R is the spectral resolution, D the telescope diameter, and Dist the stellar distance. Combining the previous three equations, we finally get that the exposure time required to reach a specified $SNR_0$ scales as:

$$t_{SNR_0} \propto \frac{R \cdot SNR_0^2 \cdot Dist^4}{\eta_{occ} \cdot Tr \cdot \eta_{PSF}^2 \cdot D^4} = \frac{R \cdot SNR_0^2 \cdot Dist^4}{\eta_p \cdot Tr \cdot \eta_{PSF} \cdot D^4} \quad , \quad (20)$$

This shows that in the exozodi-limited regime expected for observations of the most distant and time-consuming targets at visible to near infrared wavelengths, a gain in the off-axis coronagraphic point source PSF concentration ($\eta_{PSF}$) reduces exposure time faster than the overall optical transmission of the system ($\eta_{occ}.Tr$) at the planet location. It also shows that doubling the spectral resolution is less hurtful than doubling the required SNR (at least within the negligible detector noise assumption adopted), and that the exposure time required scales as $(Dist/D)^4$, as expected for point source observations limited by a spatially extended background source. In addition, the core throughput ($\eta_p$) at a given planet separation will also improve as $D$ increases, providing a further reduction in exposure time.

### 4.3 Fiducial exoearth detection

The list of nearby solar-type stars amenable to exoearth searches and spectral characterization at near-UV to near-infrared wavelengths is reasonably well known from previous studies[2,3], and an improved provisional HWO target list was recently compiled by the Exoplanet Exploration Program[101]. That preliminary list suggests that about half the anticipated number of needed HWO targets (47 "Tier A" stars) could be accessed with an IWA of 83 mas, which corresponds to an exoearth seen at quadrature around a sunlike star at 12 pc. While a significant fraction of potential HWO targets will be farther, we then adopt hereafter as a representative fiducial case a Sun/Earth system at 12 pc with an exozodiacal dust density level three times higher than in the Solar System[93].

To assess the influence of the starlight suppression KPPs (raw contrast, post-calibrated rms contrast, and off-axis core throughput), we computed the exposure times (Eq. 16) required for the (SNR=7) broadband (R=5) visible detection of a fiducial 12 pc exoearth for five combinations of raw and post calibrated contrasts, and four different core throughput values at the planet location. For the latter, we used four illustrative curves of core throughput vs. separation, meant to represent three illustrative coronagraph cases and one starshade case (Fig. 24).

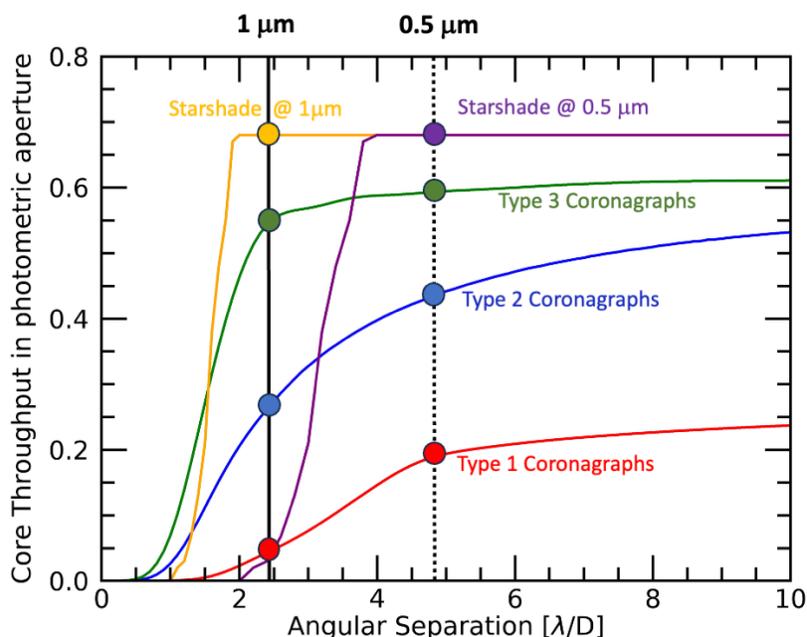

**Fig 24. Core throughput vs. separation curves adopted for the four starlight suppression systems considered. We consider three families ("types") of coronagraphs and corresponding representative core throughput curves for each. Type 1 (red curve) represents high maturity coronagraphs with low off-axis core throughput. Type 2 (blue curve) represents medium maturity coronagraphs with higher off-axis core throughput. Type 3 (green curve) represents the theoretical limit of coronagraphs and corresponds to the lowest maturity/highest core throughput case. For the notional broadband 60m starshade case considered, the core throughput only depends on *physical* separation, so that the throughput curve vs. separation expressed in λ/D units depends on wavelength (orange: 1μm, purple: 0.5 μm). The vertical black lines indicate the fiducial planet (83 mas) separation in λ/D units at 0.5 μm (dotted) and 1 μm (plain). The colored circles indicate the planet core throughput values for each starlight suppression system at 0.5 and 1 μm.**

Given that HWO's entrance pupil and overall design requirements have not yet been specified, optimized coronagraph masks have not been defined for HWO, and their core throughput vs. separation even less. To deal with that uncertainty and assess the relative merit of different options, we adopted three coronagraph types or "bins" that cover the full range of possible core throughput values and maturity levels:

- Type 1 coronagraphs represent systems with the lowest core throughput and highest technical maturity, as evidenced by the quantity and quality of available high contrast ($< 10^{-9}$) lab data (Figs. 20 & 21). This type includes the CLC, HLC, and SPC coronagraphs that will fly aboard the Roman Space Telescope[102]. All suffer from low core transmission due to pupil amplitude apodization (SPC), extended Lyot masks, and redistribution of light from the planet PSF's core into the wings because of the masks or due to the wavefront modulation imposed on the DMs as part of the dark hole solution (HLC). Such effects are especially impactful if the aperture is heavily obscured, as in the case of Roman. The core vs. separation curve adopted for Type 1 coronagraphs is the average of several APLC, SPC, and HLC coronagraph designs proposed for the USORT segmented aperture (E. Por et al. in prep, Belikov et al. in prep).

- Type 2 coronagraphs represent systems with medium-high core throughput and technical maturity. An example is the family of VVC coronagraphs, which have significantly higher throughput at small separations than Type 1 coronagraphs (e.g., Fig. 21) but are intrinsically working on a single polarization. No VVC will fly on Roman, but a VVC4 has been extensively tested in the lab on both clear and segmented apertures with measured contrast levels of a few $10^{-9}$ (Fig. 20). The core vs. separation curve adopted for Type 2 coronagraphs is the average of the VVC4 and VVC6 core throughput curves, implicitly assuming that comparable core throughput and contrast performance might be achieved on a segmented aperture with a dual polarization coronagraph[104].

- Type 3 coronagraphs represent systems with the highest possible core throughput and the lowest technical maturity. This type includes optimal N-th order coronagraphs that could theoretically be built using photonic solutions[36,105], PAPLC designs not yet tested in vacuum, and small-inner-working-angle PIAA(CMC) designs so far limited to $\sim 10^{-8}$ contrast levels in the lab. The core vs. separation curve adopted for Type 3 coronagraphs is the average of 4th, 6th, and 8th order optimal designs[36] (selected to limit sensitivity to stellar finite diameter) created for the USORT segmented aperture (E. Por et al. in prep). In addition to providing smaller inner working angles, Type 2 and 3 coronagraphs have higher throughput than Type 1 for two reasons: (i) they use masks with intrinsically higher transmission close to the optical axis; (ii) their field PSFs are very close in morphology and photometry to the non-coronagraphic ones, meaning that their PSF core fraction is close to that of a perfect Airy pattern (0.68) except near and within the inner working angle defined by the focal plane mask.

- In the starshade case, the theoretical core throughput is the highest, as no pupil mask, Lyot mask, or DM-induced PSF distortion is necessary to cancel starlight. For point sources located further out than the angle sustained by the starshade petal tip, the core throughput is set by the telescope PSF. The core vs. separation curve adopted in the starshade case is computed for a 60m-diameter starshade[106,107] flying 95 Mm in front of a 6m telescope. It is designed to provide a constant 65 mas (tip) inner working angle between $\sim 0.45$ and $\sim 1$ μm. While the laboratory starshade mask testbed reached the deepest starlight cancellation level to date, no starshade has ever flown and the technical maturity of a large space-based starshade can be considered low.

**Table 2**: List of astrophysical and instrument parameters used for broadband Earth twin detection. Parameters being varied among the different starlight suppression systems are listed in bold. C1, C2 or C3 designates coronagraphs of Type 1, 2, or 3, respectively.

| Parameter | Value |
|---|---|
| Stellar type | Solar twin |
| Stellar distance | 12 pc |
| Planet type | Earth twin |
| Planet orbit semi-major axis | 1 AU |
| Planet illumination phase and flux ratio | Quadrature; resulting in $10^{-10}$ planet-to-star flux ratio |
| Solar zodiacal light surface brightness at planet location | 23 mag/arcsec$^2$ at V band |
| Exozodiacal light surface brightness at planet location | 22 mag/arcsec$^2$ at V band for a 1 zodi solar analog |
| Exozodi Level | 3 zodis |
| Core throughput at planet location within photometric aperture ($\eta_p$) | **Varies with starlight suppression system (see Fig. 24)** |
| Core PSF fraction at planet location ($\eta_{PSF}$) | **0.4 for C1; 0.65 for C2 and C3; 0.68 for starshade** |
| (Exo)-zodi throughput at planet location (Eq 7) $\eta_{sz} = \eta_{xz} \sim \eta_{occ} = \eta_p / \eta_{PSF}$ | $\eta_p / \eta_{PSF}$ |
| Telescope (inscribed) diameter | 6m |
| Central obscuration | None |
| Central wavelength | 0.55 µm |
| Spectral resolution | 5 (i.e., 20% bandwidth) |
| End-to-end optical throughput (Tr), *excluding all starlight suppression masks and detector quantum efficiency* | **0.3 for all coronagraphs**<br>**0.5 for starshade** |
| Radius of photometric aperture | 0.7 $\lambda/D^x$ (centered at planet location) |
| Raw contrast at planet location (RC) | **$10^{-10}$ or $10^{-9}$** |
| Post calibration contrast spatial rms at the planet location ($\sigma_{\Delta C}$) | **0 or $5 \times 10^{-12}$** |
| Detector quantum efficiency (QE) | 0.9 |
| Detector noise | Negligible |
| Number of polarizations | 2 |
| Detection Signal-to-noise | 7 |

In addition to the core throughput term, the exposure time calculation includes the end-to-end optical throughput (Tr) of the system, which represents the overall transmission and reflectivity of all other ("non-coronagraphic") intervening optics. In line with previous studies[2], we assumed Tr= 0.3 for broadband imaging observations with the three coronagraph systems and Tr=0.5 with the starshade. In the starshade case, because starlight is canceled before entering the telescope, there are no DMs and no re-imaging optics are required to form additional focal and pupil planes, resulting in a simplified beam train with significantly fewer optical elements. All instrumental and astronomical parameter assumptions used in the exposure time calculations are summarized in Table 2. Note that dual polarization observations are assumed over a 20% spectral bandwidth, and that all detector noise terms are assumed to be negligible compared to the total photon noise from astrophysical background sources. In line with the SNR calculations presented in Sec 4.2., the solar and exozodi backgrounds are assumed to contribute zero mean photon noise, and ADI-like image processing is assumed, doubling the effective background noise.

---

[x] While it does not necessarily provide the optimum SNR for all starlight suppression systems and planet locations, we assume hereafter that the photometric aperture radius is 0.7 $\lambda/D$. This value maximizes SNR in the case of a point source Airy pattern to be detected against a spatially uniform background.

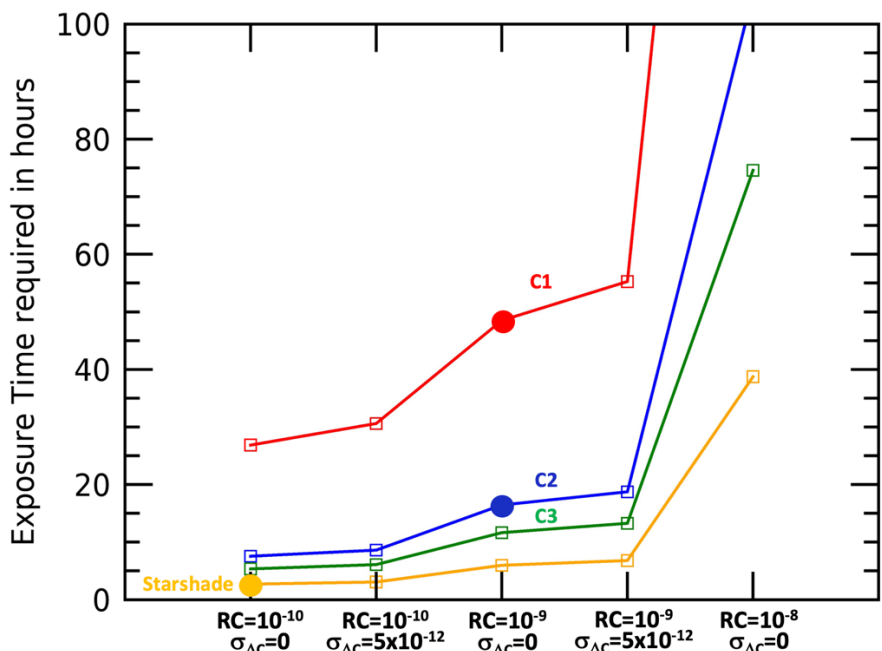

**Fig. 25. Exposure times required to detect a fiducial exoearth at 12 pc with an SNR of 7 over a 20% bandwidth centered at 0.55 μm, if various levels of raw/post-calibration contrast performance (X-axis) could be reached by four different starlight suppression systems: coronagraphs of types 1, 2, and 3 (C1, C2, C3) and starshade. For reference, the approximate raw contrast performance level currently achieved in the lab by each system type over 10–20% bandwidth is indicated by a filled circle (assuming perfect post-calibration of raw starlight speckles, i.e., $\sigma_{\Delta C}$=0). If all systems were able to achieve the same contrast performance, systems with the highest exoplanet transmission require the shortest exposures. The assumed exozodi level is 3 zodis in all cases.**

For the combination of raw contrast (RC) and post-calibration rms contrast ($\sigma_{\Delta C}$), we assumed that five different levels of performance could be achieved at the fiducial planet separation (~4.4 $\lambda$/D): (RC=$10^{-10}$, $\sigma_{\Delta C}$=0), (RC=$10^{-10}$, $\sigma_{\Delta C}$=5 x $10^{-12}$), (RC=$10^{-9}$, $\sigma_{\Delta C}$=0), (RC=$10^{-9}$, $\sigma_{\Delta C}$=5 x $10^{-12}$) and (RC=$10^{-8}$, $\sigma_{\Delta C}$=0). We note that for the detection of a fiducial exoearth with a planet-to-star flux ratio of $10^{-10}$ at an SNR of 7, a post-calibration rms contrast of ~$1.4 \times 10^{-11}$ or better is required. This common set of contrast performance values was assumed for all starlight suppression systems, regardless of the contrast levels currently demonstrated in the lab. For comparison (Sec. 2), best lab contrasts obtained to date at ~$4\lambda$/D separation range from ~$10^{-10}$ with a starshade mask to ~$10^{-9}$ for coronagraphs of Types 1 and 2, all obtained over ~10–12% bandwidth rather than the 20% assumed here, either over a half dark hole or at a single polarization.

The resulting broadband detection times for each starlight suppression type are shown in Fig. 25 for the five contrast performance levels assumed (x-axis). Absolute exposure times strongly depend on the assumptions listed in Table 2 and should be regarded as ballpark estimates. The relative exposure times between the different cases are more informative because they show the relative impact of raw contrast, post-calibration efficiency and system-specific core throughput. For a given starlight suppression system, the time to achieve a given SNR increases with raw contrast, and shoots up when the contrast goes from $10^{-9}$ to $10^{-8}$. That is even in the case of perfect speckle calibration ($\sigma_{\Delta C}$ =0), and due to prohibitive stellar shot noise. Broadband visible detection remains theoretically possible within a few hours to a day for raw contrast values up to $10^{-9}$, as long as speckles can be post-calibrated to a high accuracy (i.e., $\sigma_{\Delta C} = 5 \times 10^{-12}$ or less). However, stabilizing the raw contrast at a given level becomes more difficult as raw contrast worsens[xi].

At a fixed combination of raw contrast and post-calibration efficiency (fixed x-coordinate), the exposure time required depends only on the core off-axis throughput of each system. A significant penalty is observed for Type 1 coronagraphs compared to other systems, and the starshade system provides the shortest exposure time. However, a 60m starshade flying 95 Mm away will take weeks to travel between targets, and that time-to-SNR advantage will only hold if general astrophysics observations are conducted in the meantime. If exo-Earth blind searches and orbit determinations are required via broadband observations and revisits of many target stars, coronagraphic observations may still be preferred overall *for broadband blind detection and orbit determination of exoearths*.

### *4.4 Fiducial case for exoearth spectroscopy*

**Table 3**: List of astrophysical and instrument parameters used for Earth twin visible spectroscopy (only parameters differing from Table 2 values are listed).

| Parameter | Value |
|---|---|
| Planet type | Earth Twin with constant albedo (=0.2) |
| Solar zodiacal light surface brightness at planet location | 23 mag/arcsec$^2$ at V band. Grey opacity assumed at other wavelengths |
| Exozodiacal light surface brightness at planet location | 22 mag/arcsec$^2$ for a 1 zodi solar analog at V band. Grey opacity assumed at other wavelengths |
| Exozodi Level | 3 zodis |
| Spectral bands | Band 1: 0.45 - 0.55 µm<br>Band 2: 0.55 - 0.672 µm<br>Band 3: 0.672 - 0.822 µm<br>Band 4: 0.822 - 1.0 µm |
| Spectral resolution | 70 |
| End-to-end optical throughput (Tr), *excluding all starlight suppression masks and detector quantum efficiency* | 0.20 for IFS-based coronagraphic observations<br>0.33 for IFS-based starshade observations |
| Raw contrast at planet location (RC) | **$10^{-10}$ or $10^{-9}$** |
| Post calibration contrast spatial rms at the planet location ($\sigma_{\Delta C}$) | **0 or $5 \times 10^{-12}$** |
| Signal-to-noise per spectral bin | 10 computed at each band central wavelength, assuming continuum flux |

For spectroscopic observations, we adopt the additional instrument and astrophysical assumptions listed in Table 3. We defined four 20% spectral bands covering the full 0.45–1 µm spectral region

---

[xi] For small fluctuations, the contrast degradation caused by a given wavefront drift goes as the square root of the initial raw contrast level[35], analogous to the well-known "pinned speckles" effect.

and computed the exposure time required to reach an SNR of 10 per R=70 spectral bin at the central wavelength of each band, assuming continuum planet flux. We assumed that an integrated field spectrograph is used, resulting in lower end-to-end throughput than for broadband detection. In line with previous analysis[2], we adopted an end-to-end coronagraphic IFS throughput of 0.20, and a starshade IFS throughput of 0.33, both constant with wavelength. The detector noise is still assumed to be negligible; given the lower astrophysical flux per spectral bin and the larger number of pixels used in an IFS, this is a significantly more driving assumption than in the broadband imaging case, and one that will impact spectrometer selection and resolution.

### 4.4.1 Exposure times required for $O_2$ search

Exposure times required for spectroscopy across band 3 were computed in the middle of the band, at 0.747 μm, in the continuum near the critically important $O_2$ A band at 0.76 μm. The same five combinations of instrumental raw contrast and post processing performance are assumed as in the last section, but with reduced end-to-end throughputs for IFS spectroscopy, as defined in Table 3.

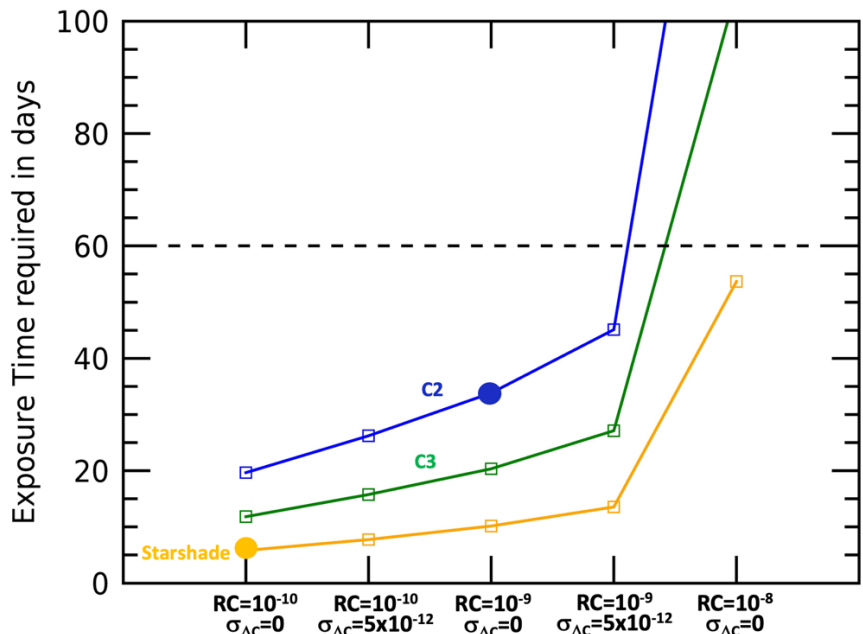

**Fig. 26. Exposure time required for R=70 spectroscopy of a fiducial exoearth at 12 pc with an SNR of 10 per spectral bin, computed at 0.747 μm, for various levels of raw/post-calibration contrast performance (X-axis) achieved by different starlight suppression systems. For reference, the approximate raw-contrast performance level currently achieved in the lab by each system is indicated by a filled circle (assuming perfect post-calibration of raw starlight speckles, i.e., $\sigma_{\Delta C}$=0). If all systems were able to achieve the same contrast performance, the starshade and Type 3 coronagraph (C3) systems would require the shortest exposures thanks to higher planet throughput. The assumed exozodi level is 3 zodis in all cases. Exposure times for Type 1 coronagraphs (not shown) are above 200 days for all contrast performance cases considered.**

As in the case of broadband detection, for the same contrast performance, the system with the highest exoplanet transmission requires the shortest exposures. In past studies[2,3], the exposure time limit for exoearth spectroscopy was set to 60 days. While somewhat arbitrary, that threshold is probably in the right ballpark given that changes in the apparent planet separation, planet illumination phase and corresponding spectrum may already be significant over such timescale, as well as possible changes in exozodi dust brightness due to resonant dust structures in nearby orbit or other time variable astrophysical phenomena. Additionally, multiplying that 60-day upper limit by the number of exoearths spectrally characterized would already result in several years of observing time. Adopting this notional cutoff limit here for simplicity, spectroscopic (R=70) measurements of a fiducial exoearth at 12 pc at an SNR of 10 per spectral bin around 0.75 μm (Fig. 26) would be within the capabilities of a starshade (in 1-2 weeks of exposure), as well as Type 2 (5-9 weeks) and Type 3 coronagraphs (2-4 weeks). That is as long as they can operate with a raw contrast of $10^{-9}$ or better and with a residual post-calibrated contrast floor ($\sigma_{\Delta C}$) of 5 x $10^{-12}$ or better, all at a separation of ~3λ/D (83 mas). For the highest maturity (Type 1) coronagraphs, the core throughput is too low to enable reasonable exposure times for spectroscopy of the fiducial target, regardless of the contrast performance achieved (exposure times >~ 200 days and off the chart). Spectroscopic observations at $10^{-8}$ raw contrast become impractically long for all coronagraphs, but may still be feasible for a starshade, assuming perfect calibration of residual starlight.

### 4.4.2 Exposure times required for spectroscopy across the full 0.45 - 1 μm range

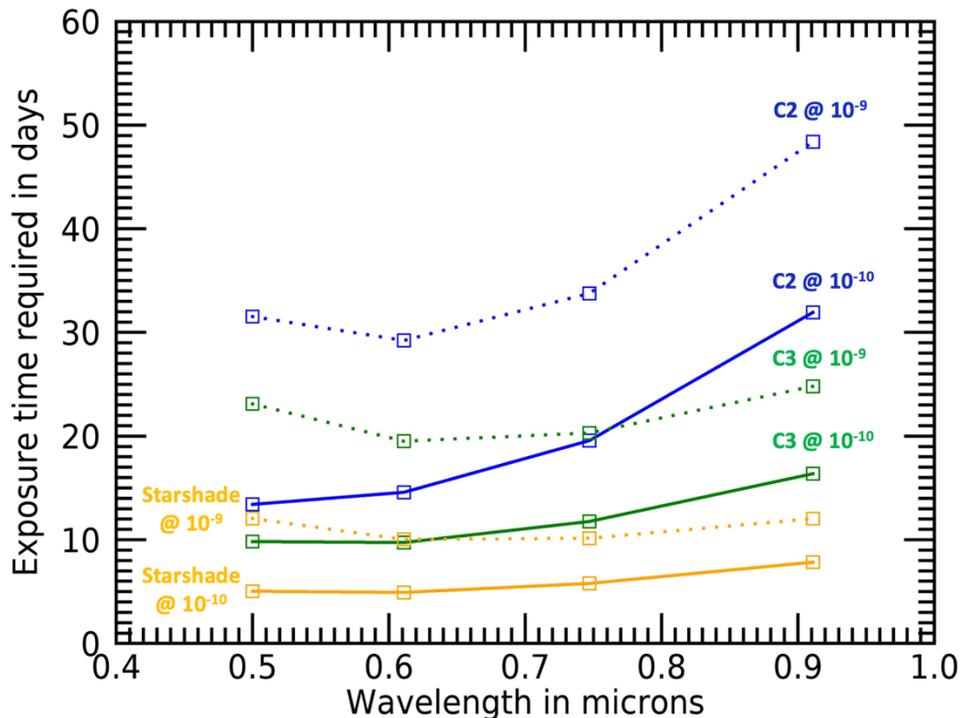

**Fig. 27. Exposure times required for R=70, SNR=10 spectroscopy of a fiducial exoearth at 12 pc as a function of optical wavelength. For each starlight suppression system considered (starshade, and coronagraphs of Types 2 and 3) the raw contrast level at the planet location is either set to $10^{-10}$ (plain curves) or $10^{-9}$ (dotted curves) All cases assume perfect speckle subtraction ($\sigma_{\Delta C}$=0). Exposure times for Type 1 coronagraphs (not shown) are around 100 days at 0.5 μm and increase sharply with wavelength.**

We now compute the exposure times required for spectroscopy as a function of visible wavelength, assuming two different raw contrast levels at all wavelengths: $10^{-10}$ or $10^{-9}$. As above, the required exposure times are calculated for SNR=10 spectroscopy per R=70 spectral bin, but now at the central wavelengths of four 20% spectral bands covering the full 0.45 μm to 1 μm range (Table 3).

The change of exposure time with wavelength differs among the starlight suppression systems considered and also depends on raw contrast performance (Fig. 27). For coronagraphs, the exposure time generally increases with visible wavelength because of two compounding effects: (i) the PSF core region area increases as $\lambda^2$, capturing more of the spatially extended (exo-)zodi background; (ii) the planet apparent separation decreases in $\lambda/D$ units and the core throughput decreases accordingly (Fig. 24). Both effects are more marked at deeper contrast, where (exo-)zodi background dominates, and for lower core throughput and PSF core fraction. This means that the required exposure times increase faster with wavelength for $10^{-10}$ contrast than $10^{-9}$, and for Type 2 coronagraphs than Type 3. For the starshade case, the core throughput is by definition (Fig. 24) the same at all wavelengths in range, and only the effect of the increased (exo-) zodi beam etendue can be seen. It results in a modest increase of exposure time with wavelength in the $10^{-10}$ contrast case, and almost a flat spectral response at $10^{-9}$ contrast.

Interestingly, and as shown in Fig. 27, exposure times can be shorter at worse raw contrast but higher throughput (e.g., a starshade operating at $10^{-9}$ contrast vs. a Type 3 coronagraph at $10^{-10}$ contrast, or a Type 3 coronagraph operating at $10^{-9}$ contrast vs. a Type 2 coronagraph at $10^{-10}$ contrast, especially for the redder wavelengths). This observation illustrates the starlight suppression system trade between core throughput and raw contrast. That is as long as raw contrast can be calibrated down to a constant level. For all starlight suppression systems considered, we further find that the spectroscopic (R=70) exposure times increase sharply at longer near infrared wavelengths, reaching 100 days or more at 1.5 μm under the combined effects of lower planet core throughput and higher exozodi background per PSF core.

Next, we investigate the time required by each system to obtain a spectrum over the full 0.45–1 μm range, either at once, or sequentially, one spectral band at a time (Fig. 28), still using the astronomical and instrumental assumptions listed in Table 3. The starshade can in principle be designed for parallel spectroscopic observations over the full 0.45–1 μm wavelength range, without a change in operating conditions. Such "full spectral parallelization" is also possible for coronagraphs but would require several coronagraph spectral channels to run in parallel over smaller individual spectral band-passes using beam-splitters or dichroics whose spectral phase and amplitude effects are controllable. Each channel would have its own pair of DMs, wavelength-dependent DM actuator settings, and coronagraphic masks. Assuming dual-polarization coronagraphic measurements over a 20% bandwidth are achievable, four parallel coronagraphs would be needed to cover the 0.45–1 μm range simultaneously.

Looking at the fiducial 12 pc exoearth case, an R=70 spectrum with an SNR of 10 per spectral bin could be obtained from 0.45 to 1 μm in 1 to 2 weeks by a starshade system, depending on the assumed contrast performance (Fig 28 top panel, orange curve). It would typically take ~4x longer, i.e., 4 to 8 weeks, depending on contrast performance, for a fully-spectrally-multiplexed Type 2 coronagraph system (blue curve), and ~2x longer, i.e. 2 to 4 weeks with a theoretically optimum coronagraph (Type 3, green curve), also composed of four parallel 20% bandwidth channels. Still considering the 6m telescope case and the fiducial exoearth, the total exposure time would be prohibitive (70–190 days) if spectra are instead obtained sequentially, 20%-bandwidth at a time with a Type 2 coronagraph (dotted blue curve), and still prohibitive (50–120 days) with Type 3 coronagraphs under most contrast performance scenarios (dotted green curve). Thus, for internal coronagraphs, the throughput and instantaneous bandwidth, per channel or synthesized via parallel channels, accessible to observations are essential factors to reigning in spectroscopic integration times to acceptable levels. To get full 0.45–1 um spectra of typical exoearths at the specified SNR (10) and spectral resolution (R=70), and if coronagraph instruments can only access 20% bandwidth at a time, some degree of spectral "multiplexing" will be required, observing in parallel with several coronagraphic channels. An important trade for conducting spectroscopic measurements with coronagraphs is between the bandwidth of individual channels, the core throughput accessible per channel and the number of channels desired for spectroscopic characterization, setting the degree of spectral parallelization and instrument complexity needed.

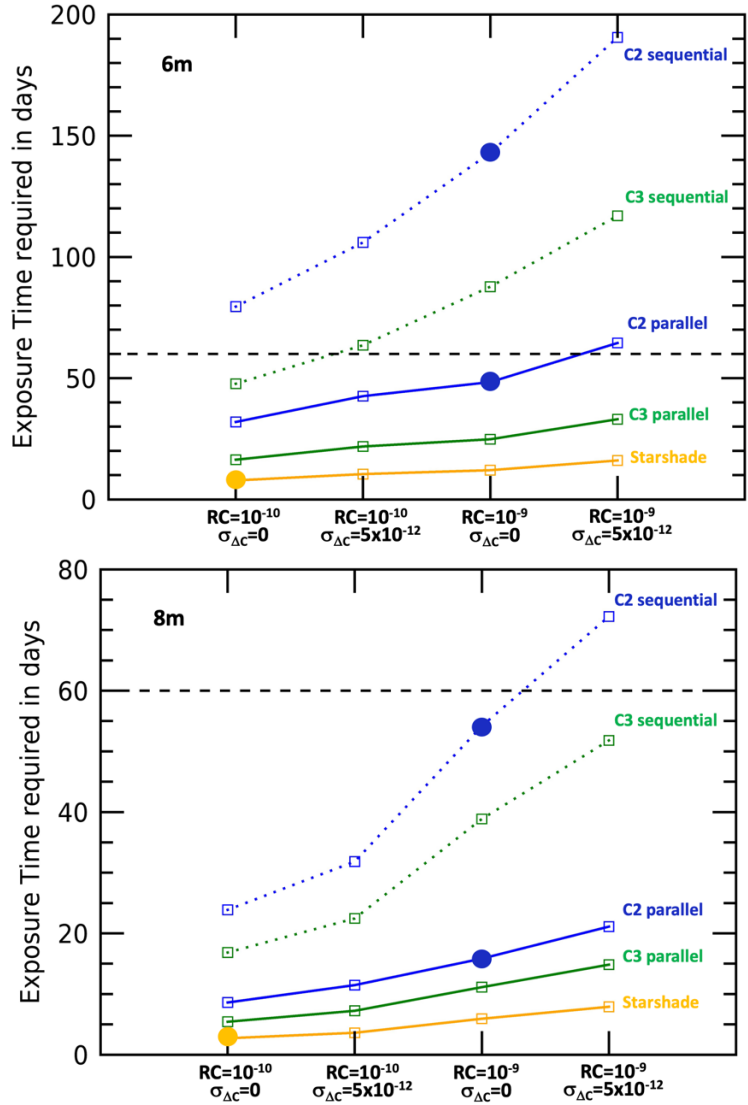

**Fig. 28. Total exposure time (in days) required to measure the R=70 SNR=10 spectrum of a fiducial 12 pc exoearth from 0.45 to 1 μm, with a notional 60-day threshold (dashed line). Five starlight suppression cases are considered: (i) an ultra-broadband (60m) starshade system covering the full wavelength range at once; (ii) four medium throughput/medium maturity coronagraph channels (type C2) working in parallel over 20% bandwidth each; (iii) a single (type C2) coronagraph channel observing sequentially over four different 20% spectral bands, one at a time. (iv) four extremely high throughput / low maturity coronagraph channels (type C3) working in parallel over 20% bandwidth each; (v) a single (type C3) coronagraph channel observing sequentially over 4 different 20% spectral bands, one at a time. For reference, the approximate raw contrast performance currently achieved in the lab by each system is indicated by a filled circle (assuming perfect post-calibration of raw starlight speckles, i.e., $\sigma_{\Delta c}$=0). Top: D=6m inscribed diameter telescope. Bottom (note the different y scale): D=8m inscribed diameter telescope, using a starshade with the same physical IWA (65 mas) as in the 6m telescope case (i.e., a ~ 62m starshade), but coronagraphs with IWAs improving as 1/D.**

In the fiducial case adopted, parallel observations using of a lower throughput coronagraph (e.g., C2 solid blue curve) can still be conducted faster than sequential observations with a higher throughput coronagraph (e.g., C3, dotted green curve). We also note that as the contrast performance gets worse, parallel observations become increasingly favored over sequential ones. It can be understood as the following: as the contrast gets e.g., from $10^{-10}$ to $10^{-9}$, observations become less limited by the (exo-)zodi background and more limited by starlight residuals. As a result, the required observing times are more uniform across spectral channels (Fig. 27), which is when parallel observations become most advantageous over sequential ones (Fig. 28).

An additional knob to turn would obviously be to use a larger telescope, as illustrated by the 8 m (inscribed diameter) case shown in the bottom panel of Fig. 28. The relative exposure times as a function of starlight suppression system throughput and level of spectral parallelization remain about the same as in the 6m case. But all exposure times go down by a factor of ~3, as expected for observations limited by shot noise from the (exo-)zodi background (exposure time scaling as $\sim 1/D^4$).

One last trade illustrated in Fig. 28 (6m panel, plain blue curve) is between raw contrast and post-calibrated contrast. For instance, for the fully parallelized type 2 coronagraphs, the exposure time required assuming $10^{-9}$ raw contrast with perfect speckle subtraction is close to that required at $10^{-10}$ raw contrast but with finite ($5 \times 10^{-12}$) post-calibrated contrast.

## 5   Design trades, numerical parametric studies, and future experiments

HWO's detailed science objectives and corresponding technical requirements remain to be defined. For direct exoplanet spectro-imaging, the basic observational parameters and quantitative figures of merit driving mission design are not yet specified. This includes the number of objects to be detected and characterized for different planet types, as well as basic measurement parameters such as wavelength coverage, spectral resolution, and signal-to-noise ratio. It is only after such high level science figures of merit, physical parameters, and measurement requirements are established that realistic mission science yield simulations can be run to optimize the observations scheduling across the pool of available target stars[31,33] and thoroughly assess the impact of individual performance parameters.

Pending such detailed mission yield simulations, we adopted in this paper a single representative exoearth target at 12pc, and used the observing time required for its detection and spectral characterization at visible wavelengths (loosely defined as 0.45–1 μm) as a simple figure of merit. The exposure time required for observations of a typical target certainly does not capture the complexity of science yield calculations over the full target sample. But it still constitutes a useful metric to assess the feasibility of scientific observations and provides valuable insight on the performance requirements of the starlight suppression system, as well as on some of the major trades remaining to be explored, numerically or experimentally.

*Numerical science yield parametric studies*

Using realistic astronomical and instrument parameters (Tables 2 & 3), we find that the time required for spectroscopy of the adopted fiducial exoearth far exceeds that needed for broadband detection and that spectral characterization drives the performance of the overall starlight suppression system. For a given spectral bandwidth, the required exposure times are set by three starlight suppression system key performance parameters (KPPs): raw contrast, post-calibrated contrast, and off-axis (planet) core throughput. The three parameters can be traded against each other to some degree, as e.g., higher off-axis throughput can make up for worse raw contrast, at least if speckles can be calibrated close to the photon noise level. Once science objectives are established, we suggest that a detailed multi-parameter trade study of science yield vs. KPP be conducted to map the combination of raw contrast, post-calibrated contrast and core throughput values that meet the science needs. This exploration can be done before implementation choices are made (e.g., trading coronagraphs with different throughputs and sensitivities to aberrations), providing a few possible winning options that can then be compared to existing lab performance, so that technical gaps and further lab work can be properly identified and prioritized.

Another important trade highlighted in our spectroscopic exposure time calculations concerns the degree of spectral parallelization, i.e., the number of individual spectral channels needed to cover the desired wavelength range. While a starshade can in principle be used for spectroscopy over the full 0.45–1 um at the same time, internal coronagraphs, if limited to 20% bandwidth per individual channel, will require multiple parallel channels to cover the same range within a reasonable time. Covering the full visible range at once could, for instance, require four parallel coronagraph channels, and twice that many if individual channels can only measure a single polarization. For internal coronagraphs, the throughput and instantaneous spectral bandwidth accessible per channel or synthesized via parallel channels, accessible to observations are hence essential factors to reigning in spectroscopic integration times to acceptable levels. A parametric study of the number of parallel coronagraph channels necessary to meet the full mission exoplanet spectroscopy yield objectives or sets of objectives reflecting different levels of ambition (e.g., spectral coverage[4,5,108-110]), would then be most useful to drive requirements and technical solutions. Broader instantaneous coronagraph bandwidth may for instance be reachable at the expense of reducing the core throughput or dark-hole field of view.

Pending such detailed parameter studies, we find (Fig. 28) that full visible-range spectroscopy of an exoearth orbiting a Sunlike star at the typical distance of HWO targets (taken to be 12 pc) with a 6m telescope is possible for starlight suppression systems providing a raw contrast of $10^{-9}$ or better with a residual post-calibrated contrast floor ($\sigma_{\Delta C}$) of 5 x 10$^{-12}$ or better, and a core throughput greater than ~ 30%, all at a separation of <~3λ/D.

*Technical demonstrations: coronagraphs*

In comparison, the best coronagraphic lab results so far (Sec. 2, Fig. 20) have reached the required level of raw contrast at 3–4 λ/D separation but only over a 10% spectral bandwidth (CLC & HLC lab set-ups), with < 15% off-axis throughput and only on a monolithic unobscured aperture. The VVC4 10% bandwidth set-up demonstrated raw contrast values a few times worse at significantly higher core throughput (Fig. 21), but still with a monolithic aperture and using a single polarization.

For segmented off-axis apertures, the best polychromatic performance currently achieved is about 10x worse, with ~10$^{-8}$ raw contrast demonstrated at 3 λ/D. However, among the coronagraph types providing the best clear aperture results, only one (VVC4) was also tested in vacuum in that off-axis segmented configuration and using a single DM instead of two. This makes it currently difficult to isolate the effect of segmentation on coronagraphic performance. In addition, we caution that the current vacuum "segmented" experiments used a segmented aperture mask where only the amplitude shows discontinuities across the pupil, not the phase. Such experiments hence demonstrate the contrast limit accessible with perfectly co-phased segments, but do not realistically probe the contrast sensitivity to individual segments tip-tilt and piston errors. Given that HWO is currently baselining a segmented telescope, this highlights the pressing need for more vacuum lab demonstrations of high contrast coronagraphic imaging and spectroscopy on segmented apertures, especially with entrance pupils that exhibit both phase and amplitude discontinuities. Nonetheless, experiments on clear monolithic apertures should also still be pursued because they provide the adequate benchmark for assessing the intrinsic impact of segmentation. To provide the most direct comparison between the monolithic and segmented cases, it is desirable that the same set-up be used, limiting changes to what is strictly required to deal with segmentation (e.g., mask adjustments but same operating conditions). Further vacuum tests on monolithic apertures will also help us to understand and improve the current contrast limits set by the testbeds themselves, and by the residual defects of manufactured masks. In most cases, it is inferred that mask defects are responsible for a significant part of the contrast performance degradation observed as spectral bandwidth increases. This highlights the need for improved mask manufacturing, as well as detailed mask inspection and characterization *before* coronagraphic testing. When incorporated in WF control algorithms, the precise knowledge of residual defects in manufactured masks may indeed result in faster dark hole convergence and improved contrast performance.

An additional degradation of broadband contrast performance is currently observed in the laboratory when switching from off-axis to on-axis segmented apertures, for which the best raw contrast achieved at 3 λ/D is ~ 7 x 10$^{-8}$, about 100x worse than in the off-axis monolithic case. However, that degradation is far larger than predicted by current models and again based on vacuum testing of a single coronagraph (PIAACMC). This calls for further laboratory testing in the next few years so that any firm conclusion on the gap currently observed in the lab between off- and on-axis coronagraphic results can be reached in a timely manner.

Given the extreme level of contrast (less than about $10^{-9}$) required for direct exoearth spectroscopic observations, laboratory experiments have mainly concentrated so far on demonstrating deep broadband (>~10%) raw contrast at small angular separations. However, once the contrast performance is good enough that residual starlight levels fall below the signal from irreducible astrophysical background sources (zodi and exozodi dust), the exposure time required to characterize exoearth planets becomes essentially independent of raw contrast[xii]. Consequently, the predicted total mission yield of exo-Earths spectrally characterized increases only marginally with further raw contrast improvements[32]. Depending on the stellar distance and exact *visible* wavelength, that transition occurs at a threshold raw contrast level ranging from ~$10^{-10}$ to ~$10^{-9}$ (Sec. 4.1). Once that raw contrast threshold performance is reached, the ability to spectrally characterize exoearths primarily depends on the next two starlight suppression system KPPs: off-axis core throughput, especially at small angular separations; and post-calibrated contrast achieved after image processing. Both parameters have been fairly overlooked so far and deserve a lot more attention going forward.

As far as throughput is concerned, accessing high core throughput at separation as close as 2-3 $\lambda/D$ will be mandatory for successful spectroscopic observations of exoearths with HWO. There are three throughput components to maximize: the overall end-to-end optical transmission of the telescope + instrument system, excluding all starlight suppression optics; the cumulative transmission of all occulting masks; and the fraction of planet light captured in the off-axis PSF core. The first term applies equally to all sources in the field, while the latter two are field-dependent. Because observations of the most distant and time-consuming targets are expected to be limited by the spatially extended (exo)-zodi backgrounds, a special emphasis should be put on maximizing the PSF core fraction at the planet location rather than the other two terms (Sec. 4.2, Eq. 20). A trend currently observed is that the raw contrast level demonstrated by coronagraph set-ups appears to degrade as their throughput at small separations increases (Fig 20 & 21). Because coronagraphs with higher throughput at close-in separations also tend to exhibit higher sensitivity to low-order aberration drifts (Por et al. in prep), this would be expected for dynamic contrast performance, but not in the lab under essentially static conditions. Whether this unpredicted trend is real or due to small number statistics remains to be explored.

The least experimentally constrained performance parameter, at least for contrast levels below $10^{-7}$, is the post-calibrated contrast rms error ($\sigma_{\Delta C}$), achieved after estimation and subtraction of starlight speckles using differential imaging techniques. However, this parameter has a crucial impact on the feasibility of exoearth spectroscopic observations, which will require $\sigma_{\Delta C}$ values or ~$10^{-11}$ or lower. Additionally, if post-calibration speckle residuals can be kept below the (exo)-zodi background shot noise level, the raw contrast performance could be relaxed to the threshold value at which (exo)-zodi background dominates. Given the existing trade between raw contrast and off-axis core throughput, this would open the possibility to operate at higher core throughput values than commonly thought, a very attractive prospect.

---

[xii] At least if post-calibration of residual speckles can be achieved down to rms levels lower than the (exo-)zodi background shot noise.

An even more crucial open question is whether information provided contemporaneously with the science images (rather than through conventional asynchronous ADI/RDI), using either WFS data or data recorded outside the dark hole (spatially or spectrally), could yield precise estimates of starlight residuals and of their uncorrected temporal fluctuations. The latter would represent a real paradigm shift, from a wavefront stability demand to a wavefront knowledge demand. Wavefront stability would still be required to maintain the average raw contrast below the background limited threshold level, but no longer within a fraction of the planetary signal. As in the raw contrast case, tolerating higher levels of wavefront and raw contrast fluctuations could have a tremendous impact on fundamental design trades such as telescope stability vs. coronagraph resilience to aberrations ("robustness"), and coronagraph robustness vs. core raw throughput. Diving into such trades is beyond the scope of this paper but is the object of on-going community work (Por et al. in prep). Given the importance of the post-calibrated contrast parameter for the design of HWO's starlight suppression system, substantial efforts should be spent to minimize it and estimate it, e.g. through dedicated coronagraphic lab experiments or by using science + telemetry data from space coronagraph instruments on-board JWST and ultimately on Roman[79].

The ability to accurately subtract residual speckles may also depend heavily on the concepts of operations and observing scenario. Exploring the overall system trade between observatory stability, the coronagraph WFS/C performance, and image processing contrast-improvement capabilities is a very high priority to allow definition of HWO's coronagraph requirements for the detection and spectroscopy of exoplanets, in particular for the driving science case of exoearth spectroscopic observations. The observatory and its coronagraph instruments (including their WFS/C system, operations concept and post-processing strategy) should be jointly optimized as a single system, so that the requirement allocations may be properly split between the telescope and the coronagraph systems. Making the necessary design trades will require a detailed error budget, informed by end-to-end integrated modeling of all the different sub-system components as well as model validations of coronagraph testbed results and post-processing improvements, reaching all the way to projected multi-variate exoplanet science yields and molecular abundance spectral retrievals.

*Technical demonstrations: starshades*

The current baseline for exoplanet direct broadband detection with HWO is a coronagraphic system capable of conducting agile blind searches and orbital characterization of exoplanets via repeated observations at low spectral resolution (R~5). Unless it can be refueled through in-orbit servicing, a starshade will be limited in its number of target slews and not so well suited to a blind exoearth search phase[111]. However, a starshade holds many promises for the spectral characterization of exoplanets and exoearths previously identified by HWO's coronagraph or other precursor observations: this includes deep contrast at small separations over large (~ 100%) instantaneous spectral bandwidths at high off-axis (planet) core throughput. Also, because it blocks starlight before entering the telescope, a starshade system works equally well with off-axis, on-axis, monolithic or segmented apertures. As long as the telescope remains diffraction limited, no sub-nm wavefront stabilization system, additional starlight suppression masks, re-imaging optics or deformable mirrors are required in the beam train. As a result, the overall system throughput is intrinsically high, the native telescope PSF is retained for off-axis sources, and the dark hole angular size is only limited by the detector size rather than by the number of DM actuators. Some of these promises have been confirmed in the lab, with a small-scale starshade mask experiment already reaching $10^{-10}$ contrast over a 12% bandwidth down to a separation of only $2\lambda/D$, with starlight residuals falling off quickly at wider separations.

These impressive small-scale results confirm the predictions of vectorial finite diffraction theory, and already demonstrate a level of starlight suppression performance commensurate with the detection and spectroscopy of exoearths. They also show that the challenges of operating a starshade on a large telescope are elsewhere. Indeed, in the starshade case, the wavefront control and starlight suppression challenges are essentially shifted from the telescope and internal WFS/C systems to deploying a large external occulter with high shape accuracy and stability, flying in precise formation with the telescope to block the starlight, and to mitigating scattered sunlight. NASA's starshade technology development ("S5") activity[112,113] has closed several milestones in these areas and in particular has closed two major technology gaps for formation flying sensing and starlight suppression/model validation[65]. Milestones pertaining to the remaining technology gap, i.e., deployment accuracy and shape stability, are expected to be closed in FY24 and will concentrate on extending the currently demonstrated mechanical performance to higher-fidelity components.

Given the current lab results, as well as the potential for reaching deep broadband contrast at small separations and high throughput, it appears appropriate to keep the starshade approach in HWO's toolbox. When working in tandem with the HWO coronagraphic system, a starshade could significantly enhance spectroscopic performance in the visible and extend it to the near infrared for many targets, thanks to its small inner working angle. It may also offer a unique solution for photon-efficient exoplanet observations with HWO in the near UV, should the science community recommend them. A starshade could be launched with the HWO prime mission or a few years later, after the coronagraph has identified which stars have exoearth candidates orbiting in their habitable zones.

If such a dual coronagraph plus starshade option were to be pursued, further starshade technology maturation would be required to test for manufacturing accuracy and thermal stability after full-scale petal development of an HWO-compatible starshade, which could be of order 60 m in diameter at visible/near infrared wavelengths or ~35m diameter for near-UV observations. Additionally, the starshade approach cannot be fully tested at scale from the ground and there is no technology demonstration mission like the Roman coronagraph currently planned in the coming years. This raises the question of whether a starshade space demonstration is opportune and necessary in the HWO context, and what its minimum size should be to remain technically relevant.

*Summary*

The exoearth spectral characterization objectives of HWO will drive HWO's design and remain to be defined. This includes the range of insolation levels and sizes to be adopted for exoearths, the exact number to be characterized (or searched for at some completeness level), the wavelength range to be accessed, the signal-to-noise ratio and spectral resolution desired, as well as the distribution of such observing parameters across the sample. As an illustrative scenario, we examined here the case of an Earth twin orbiting a sunlike star at 12 pc with three times the level of solar zodiacal dust (our current best estimate) and required that a full spectrum extending from 0.45 to 1 μm be obtained at an SNR of 10 and spectral resolution of 70.

Assuming such broad spectral characterization is desired, we identify three main options for conducting time-efficient spectroscopy of a significant number of exoearths at visible wavelengths with HWO: (i) improved broadband coronagraph systems, likely using several parallel spectral channels, that provide deep raw ($< 10^{-9}$) and post-calibrated ($< 5 \times 10^{-12}$) contrasts together with high off-axis (exoplanet) throughput at $<\sim 3\lambda/D$ separation; (ii) use of a starshade for photon-efficient enhanced broadband spectroscopy of exoearths and other planets previously detected by HWO's coronagraph or by other precursor indirect observations (e.g, extreme precision radial velocimetry); (iii) use of a larger (e.g., 8m) telescope to significantly reduce exposure time in the (exo-)zodi limited regime expected for many targets, especially the most distant and time-consuming ones. Any of these options would strongly benefit HWO's exoearths spectral characterization capabilities. They could also be exercised in conjunction for optimum performance.

Finally, extending HWO's exoearth spectroscopic measurements to the near UV, where planets are intrinsically faint in reflected light, or to the near infrared, where planets are less resolved from their parent star and the irreducible (exo-)zodi background signals are higher than at visible wavelengths, would make these options even more indispensable.

Our findings may provide useful input to the on-going design trade space exploration conducted independently by the HWO Science, Technology and Architecture Review Team (START) and Technology Assessment Group (TAG).

*Code and Data Availability*

NASA regulations govern the release of source code, including what can be released and how it is made available. Readers should contact the corresponding author if they would like copies of the software or data produced for this study.

*Acknowledgments*

Part of this research was carried out at the Jet Propulsion Laboratory, California Institute of Technology, under a contract with the National Aeronautics and Space Administration.

**First Author** is a principal research scientist at the Jet Propulsion Laboratory, California Institute of Technology. He received his MS degree in physical engineering from the Ecole des Mines (France) in 1993, as well as MS and PhD degrees in astrophysics and space techniques from the University of Paris VII in 1994 and 1999, respectively. His expertise is in the design, assembly and scientific exploitation of high-contrast high-resolution optical systems – coronagraphs, interferometers and starshades – used for astronomical imaging. His main scientific focus is on the direct detection and spectroscopy of exoplanetary systems, ultimately leading to the search for life on Earth-like exoplanets through the use of new instruments and data reduction techniques. Dr Mennesson is the author or co-author of over 300 publications in specialized journals and at professional conferences.